\documentstyle[12pt]{article}
\def\journal#1, #2, #3, #4 { {\sl #1~}{\bf #2~}(#3) #4 }

\def\mpl{\journal Mod. Phys. Lett., }

\def\pr{\journal Phys. Rev., }

\def\prl{\journal Phys. Rev. Lett., }

\def\ptp{\journal Prog. Theor. Phys., }

\def\cmp{\journal Comm. Math. Phys., }

\def\np{\journal Nucl. Phys., }

\def\pl{\journal Phys. Lett., }

\catcode`\@=11
\def\marginnote#1{}
\newcount\hour
\newcount\minute
\newtoks\amorpm
\hour=\time\divide\hour by60
\minute=\time{\multiply\hour by60 \global\advance\minute
by-\hour}\edef\standardtime{{\ifnum\hour<12
\global\amorpm={am}%
        \else\global\amorpm={pm}\advance\hour by-12 \fi
        \ifnum\hour=0 \hour=12 \fi
        \number\hour:\ifnum\minute<10
0\fi\number\minute\the\amorpm}}
\edef\militarytime{\number\hour:\ifnum\minute<10
0\fi\number\minute}

\def\draftlabel#1{{\@bsphack\if@filesw {\let\thepage\relax
   \xdef\@gtempa{\write\@auxout{\string
      \newlabel{#1}{{\@currentlabel}{\thepage}}}}}\@gtempa
   \if@nobreak \ifvmode\nobreak\fi\fi\fi\@esphack}
        \gdef\@eqnlabel{#1}}
\def\@eqnlabel{}
\def\@vacuum{}
\def\draftmarginnote#1{\marginpar{\raggedright\scriptsize\tt#1}}
\def\draft{\oddsidemargin -.5truein
        \def\@oddfoot{\sl preliminary draft \hfil
        \rm\thepage\hfil\sl\today\quad\militarytime}
        \let\@evenfoot\@oddfoot \overfullrule 3pt
        \let\label=\draftlabel
        \let\marginnote=\draftmarginnote

\def\@eqnnum{(\theequation)\rlap{\kern\marginparsep\tt\@eqnlabel}%
\global\let\@eqnlabel\@vacuum}  }

\def\numberbysection{\@addtoreset{equation}{section}
        \def\theequation{\thesection.\arabic{equation}}}

\def\underline#1{\relax\ifmmode\@@underline#1\else
        $\@@underline{\hbox{#1}}$\relax\fi}

\catcode`@=12
\relax

\numberbysection
\def\beq{\begin{equation}}
\def\eeq{\end{equation}}
\def\beqa{\begin{eqnarray}}
\def\eeqa{\end{eqnarray}}
 \def\nnn{\nonumber \\}
\def\sqr#1#2{{\vcenter{\vbox{\hrule height.#2pt
\hbox{\vrule width.#2pt height#1pt \kern#1pt
\vrule width.#2pt}
\hrule height.#2pt}}}}

\def\lfloorhat{{\hat \lfloor}}
\def\rfloorhat{{\hat \rfloor}}
\def\lceilhat{{\hat \lceil}}
\def\rceilhat{{\hat \rceil}}

\def\hhat{{\widehat h}}

\def\Jhat{{\widehat J}}

\def\dhat{{\widehat d}}
\def\chat{{\widehat c}}
\def\Cp{ C'\, }
\def\cp{ c'\, }
\def\Chat{{\widehat C}}
\def\Chatp{{\widehat C}'\,}
\def\Dp{ D'\, }
\def\Dhat{{\widehat D}}
\def\Dhatp{{\widehat D}'\,}

\def\Deltahat{{\widehat \Delta}}
\def\mhat{{\widehat m}}
\def\nhat{{\widehat n}}

\def\psip{ \psi'\, }
\def\psihat{{\widehat \psi}}

\def\psib{{\overline \psi}}

\def\psibhat{{\widehat{\overline \psi}}}

\def\Mhat{{\widehat M}}
\def\Mhatp{{\widehat M}'\, }

\def\xihat{{\widehat \xi}}

\def\xib{{\overline \xi}}

\def\xibhat{{\widehat{\overline \xi}}}

\def\muhat{{\widehat \mu}}
\def\nuhat{{\widehat \nu}}
\def\lambdap{ \lambda'\, }
\def\lambdahatp{{\widehat \lambda}'\, }
\def\lambdahat{{\widehat \lambda}}

\def\rhat{{\widehat r}}
\def\shat{{\widehat s}}
\def\that{{\widehat t}}
\def\qhat{{\widehat q}}
\def\varpip{\varpi'\,}
\def\varpihat{{\widehat \varpi}}
\def\varpihatp{{\widehat \varpi}'\,}
\def\varpib{{\overline \varpi}}

\def\mhat{{\widehat m}}
\def\nhat{{\widehat n}}

\def\square{\mathchoice\sqr34\sqr34\sqr{2.1}3\sqr{1.5}3}
\def\qed{ $ \quad\square $ }

\begin{document}

\begin{titlepage}

\nopagebreak
\begin{flushright}

LPTENS--91/22, \\
hep-th@xxx/9205034 \\
                April  1992
\end{flushright}

\vglue 1.5  true cm
\begin{center}
{\large\bf
 GRAVITY-MATTER COUPLINGS\\
 FROM LIOUVILLE THEORY}
\vglue 1.5 true cm
{\bf Jean-Loup~GERVAIS}\\
{\footnotesize Laboratoire de Physique Th\'eorique de
l'\'Ecole Normale Sup\'erieure\footnote{Unit\'e Propre du
Centre National de la Recherche Scientifique,
associ\'ee \`a l'\'Ecole Normale Sup\'erieure et \`a
l'Universit\'e
de Paris-Sud.},\\
24 rue Lhomond, 75231 Paris CEDEX 05, ~France}.
\end{center}
\vfill
\begin{abstract}
\baselineskip .4 true cm
\noindent
The three-point functions   for minimal models  coupled to gravity
 are derived in the operator approach
to Liouville theory which is  based on its  $U_q(sl(2))$ quantum group
structure.  The gravity-matter
coupling is formulated  by treating the  latter as a continuation
of the former. The result is  very simple, and shown to agree
with matrix-model calculations  on the
sphere. The precise definition of the corresponding
cosmological constant is
given
in the operator  solution of the quantum Liouville theory.
It is shown that  the symmetry between
quantum-group spins $J$ and $-J-1$ previously
put forward by the author
 is the explanation of the continuation in the number of
screening operators discovered by Goulian and Li.
Contrary to the previous discussions of this problem, the present
approach clearly separates the emission operators for each leg.
This  clarifies the structure of the dressing by gravity. It is shown,
in particular that the end points are not treated on the same footing
as the mid point. Since the outcome is completely symmetric this
suggests  the possibility of a picture-changing mechanism.

\end{abstract}
\vfill
\noindent {\footnotesize  typeset by \it Paul-Serge J. Ianov}

\end{titlepage}
\baselineskip .5 true cm
\section{Introduction  }
\markboth{ 1. Introduction }{ 1. Introduction }
\label{1}

Matrix models have shown that  minimal models become much
simpler when they are coupled
to 2D gravity.  This was  understood\cite{GL,D,Coulomb}
in the Coulomb-gas
formulation by  performing a  continuation to
negative numbers of screening operators. It is the purpose
of the present article to study  this phenomenon in
the operator  treatment of the Liouville theory
which is based on   the quantum group structure
displayed\cite{B,G1,G2,G3,G4,GR,CG},
in recent years. The results discussed here
 were  already reported some time ago\cite{G5},
in an abbreviated form.

It should be stressed that the tools
needed for deriving the three-point function are  already
present in  earlier papers of mine\cite{G1,G2,G3,G4}.
In particular, the building of operators with positive
weights  in the strong-coupling regime led to the unravelling
of a symmetry between quantum group spins $J$ and $-J-1$
some time  ago\cite{G2,G3}. This symmetry is basically why
the Goulian and Li continuation\cite{GL} is valid. Moreover,
the construction of a local Liouville
field was already done before\cite{G4}.
 Finally, the universal chiral family
associated with $U_q(sl(2))$ that comes out by quantizing  Liouville
theory, was completely determined in \cite{G1}, including the
detailed
normalisation of the operators, so that their matrix elements
between highest-weight states may be read off from \cite{G1}.
This is enough to determine the three-point functions, as we shall
show below.

The outcome of the present study is threefold. First,
 we shall spell out  the drastic
simplifications that occur in the present approach
when gravity and matter are coupled.
This is useful,  in particular,  since
 the quantum group structure mentioned
above\cite{G1,G2,G3,G4,GR,CG},
gives  general
formulae  which remain rather
complicated, compact as they may be.
Second, the other  discussions\cite{GL,D,Coulomb}
deal with each correlation function  separately, and thus remain at the level
of conformal blocks, where the screening operators are not
specifically attributed to any particular external leg.
Thus they do not clarify how  world-sheet operators
such as powers of the Liouville field are  given as
 sums over  products of holomorphic and antiholomorphic
vertex operators   with  specified
screening charges. Expanding ref.\cite{G4}, we shall first
(sections  2, and 3)
construct the nedeed operators once for all by imposing
the physical requirement that the exponentials
of the Liouville field be local and  possess a consistent
restriction to the space of states with equal right and left
momenta. Then, (sections  4, and 5) we shall show   that
appropriate  matrix elements agree with the result of matrix-models.
The operator content of the world-sheet fields is  clear
in the present discussion, contary to the Coulomb gas picture.
In particular we shall  see  that,  in the three-point function,  the
endpoints are treated on a different footing even if the result is
completely symmetric.
Last,  we shall show how the cosmological constant should be
introduced in the present scheme. This will give us an operator
derivation of the DDK argument\cite{Da,DK}, which may be
immediately applied on the sphere and on the torus (see section 5).

In order to set the stage, let us first  recall some feature of the
classical  Liouville dynamics. In the conformal gauge,
it  is governed by the action:
\begin{equation}
 {\it S}={1\over 4\pi } \int d_2x
\sqrt{\widehat g} \Bigl\{ {1\over 2} {\widehat
g}^{ab}\partial_a\Phi
\partial_b\Phi+
e^{\displaystyle 2\sqrt{\gamma} \Phi}
+{1\over 2\sqrt{\gamma}}R_0 \Phi\Bigr \}
\label{1.1}
\end{equation}
 ${\widehat g}_{ab}$ is the fixed background metric. We work
for fixed genus, and do not integrate over the moduli. As  is
well known,  one can choose a local coordinate system such that
${\widehat
g}_{ab}=\delta_{a b}$.
 Thus we are reduced to the action
\begin{equation}
 {\it S}={1\over 4\pi } \int d\sigma d\tau
\, \Bigl ( {1\over 2}({\partial \Phi\over \partial \sigma})^2
+{1\over 2}({\partial \Phi\over \partial \tau})^2
+e^{\displaystyle 2\sqrt{\gamma} \Phi} \Bigr )
\label{1.2}
\end{equation}
where $\sigma$ and $\tau$ are the local coordinates.  The
complex structure is assumed to be such that the curves  with
constant $\sigma$ and $\tau$ are everywhere
tangent to the local imaginary
and real axis  respectively.
In a typical situation, one may  work on the cylinder $0\leq
\sigma \leq 2\pi$,  $-\infty \leq \tau \leq \infty $ obtained by
an
appropriate mapping from one of the handles of a general
Riemann surface, and we shall do so in the present article.
The action \ref{1.2}  corresponds to a conformal
 theory such that $\exp(2\sqrt{\gamma} \Phi) d\sigma d\tau $
is invariant.
  The classical equivalent of the chiral vertex operators
 may be obtained  very simply\cite{GN1,GN2,G4} by   using the fact that
the field $\Phi (\sigma,\, \tau)$
satisfies the equation
\begin{equation}
 {\partial^2 \Phi\over \partial \sigma^2}
+{\partial^2 \Phi\over \partial \tau^2}=
2 \sqrt{\gamma}\> e^{\displaystyle 2\sqrt{\gamma}\Phi}
\label{1.3}
\end{equation}
if and only  if
\begin{equation}
e^{-\displaystyle \sqrt{\gamma} \Phi}={i\sqrt{\gamma \over 2}}
\sum_{j=1,2} f_j(x_+)
g_j(x_-);
  \quad  x_\pm=\sigma\mp i\tau
\label{1.4}
\end{equation}
where $f_j$ (resp.($g_j$), which are functions of a single
variable,  are
solutions of the  same Schr\"odinger equation
\begin{equation}
-f_j''+T(x_+)f_j=0,\quad
\hbox{( resp.}\>  -g''_j+\overline T(x_-)g_j\,\hbox{)}.
\label{1.5}
\end{equation}
 The solutions
are normalized so that their Wronskians $f_1'f_2-f_1f_2'$
and $g_1'g_2-g_1g_2'$ are equal to one.
The proof of this basic fact is straightforward
\cite{GN1,GN2,G4}. The potentials
$T(x_+)$ and $\overline T(x_-)$ are the two
components of the stress-energy tensor, and, after quantization,
Eqs.\ref{1.5}  become the  Virasoro Ward-identities
 associated with  the vanishing of the singular vector
at the second level. As a result the Liouville theory also
describes minimal models provided the coupling constant
$\gamma$ is taken to be negative. This is how, we shall treat
the matter fields.
 For the dynamics associated with the
action Eq.\ref{1.2}, $\tau$ is the time variable, and the canonical
Poisson
brackets are
\begin{equation}
\bigl \{\Phi(\sigma_1, \tau),
{\partial\over \partial \tau}  \Phi(\sigma_2, \tau) \bigr
\}_{\hbox{P.B.}}
=4\pi \,  \delta(\sigma_1-\sigma_2),\quad
\bigl \{\Phi(\sigma_1, \tau),\Phi(\sigma_2, \tau) \bigr
\}_{\hbox{P.B.}}= 0
\label{1.6}
\end{equation}
The cylinder  $0\leq
\sigma \leq 2\pi$,  $-\infty \leq \tau \leq \infty$
 may be mapped on the complex
plane of $z=e^{\tau+i\sigma}$, and the above Poisson brackets
lead to the usual radial quantization.

A priori, any two
pairs
$f_j$ and $g_j$ of linearly independent solutions of Eq.\ref{1.5}
are suitable. In this connection, it is convenient to rename the
functions $g_j$ by letting $\overline f_1=- g_2$,
$\overline f_2= g_1$.
 Then   one  easily sees  that Eq.\ref{1.4} is
left unchanged if $f_j$ and $\overline f_j$ are replaced by
$\sum_kM_{jk}f_k$ and   $\sum_kM_{kj}\overline f_k$,
respectively, where
$M_{jk}$ is an arbitrary constant matrix with
 determinant  equal to one.
Eq.\ref{1.4} is
$sl(2,C)$-invariant with $f_j$ transforming as a representation of
spin $1/2$. At the quantum level, the $f_j$'s and $\overline f_j$'s
 become operators that do not commute, and  the
group $sl(2)$ is deformed to become  the quantum group
$U_q(sl(2))$.
 This  structure plays a crucial role at
 the quantum level, and we now elaborate
upon  the classical  $sl(2)$ structure  where the calculations are simple.

At the classical level, it is trivial to take Eq.\ref{1.4} to any
power.  For positive integer  powers $2J$, one gets
(letting $\beta=i\sqrt {\gamma \over 2}$)
\beq
e^{-\displaystyle 2J \sqrt{\gamma} \Phi}=
\sum_{M=-J}^J {\beta^{2J}(-1)^{J-M} (2J) !\over (J+M) !(J-M) !}
\left (f_1(x_+)\,\overline f_2(x_-)\right )^{J-M}
\left(f_2(x_+) \overline f_1(x_-)\right )^{J+M}.
\label{1.7}
\eeq
It is convenient  to put the result under the form
\begin{equation}
e^{-\displaystyle 2J \sqrt{\gamma} \Phi}=
\beta^{2J}
\sum_{M=-J}^J (-1)^{J-M}
f_M^{(J)}(x_+) \overline f_{-M}^{(J)}(x_-).
\label{1.9}
\end{equation}
where $J\pm M$ run over integer. The $sl(2)$-structure has been  made
transparent by letting
\beq
f_M^{(J)}\equiv \sqrt { \textstyle {2J\choose J+M}}
\left (f_1\right )^{J-M} \left(f_2\right )^{J+M},  \quad
\overline f_M^{(J)}\equiv\sqrt { \textstyle {2J\choose J+M}}
\left (\overline f_1\right )^{J+M}
\left(\overline f_2\right )^{J-M}.
\label{1.8}
\eeq
The notation anticipates that $f_M^{(J)}$ and $\overline f_M^{(J)}$
form   representations of spin $J$. This is  indeed true since
$f_1$, $f_2$ and $\overline f_1$, $\overline f_2$    span
spin $1/2$ representations, by construction.   Explicitely one finds
$$I_\pm f_M^{(J)}=\sqrt {(J\mp M)(J\pm M+1)} f_{M\pm 1}^{(J)},
\quad
I_3 f_M^{(J)} =Mf_M^{(J)}
$$
\begin{equation}
\overline I_\pm \overline f_M^{(J)}
=\sqrt {(J\mp M)(J\pm M+1)} \overline f_{M\pm 1}^{(J)}, \quad
\overline I_3 f_M^{(J)} =M\overline f_M^{(J)},
\label{1.12}
\end{equation}
where $ I_\ell$ and $\overline I_\ell$ are the
infinitesimal generators of the   $x_+$ and  $x_-$ components
respectively. Moreover, one sees that
\begin{equation}
\left (I_\ell+ \overline I_\ell\right )
e^{-\displaystyle 2J \sqrt{\gamma} \Phi}=0
\label{1.13}
\end{equation}
so that the exponential of the Liouville field are group
invariants.

At this simple classical level one may continue trivially to
negative spins. This gives exponentials  of $\Phi$ with a
positive exponent, and in particular defines the cosmological term
$e^{\displaystyle 2  \sqrt{\gamma} \Phi}$  which is
the potential term of the action Eq.\ref{1.2}. These terms may be
written in  two
equivalent ways:
\begin{equation}
e^{\displaystyle 2J \sqrt{\gamma} \Phi}=
\beta^{-2J}
\sum_{M>J,\,   \hbox{\scriptsize or} M<J}
f_M^{(-J)}(x_+) \overline f_{-M}^{(-J)}(x_-).
\label{1.14}
\end{equation}
where
$$f_M^{(-J)}\equiv \sqrt { \textstyle (-1)^{J+M}{-2J\choose -J+M}}
\left (f_1\right )^{J-M} \left(f_2\right )^{J+M},
$$
\beq
\overline f_M^{(-J)}\equiv\sqrt { \textstyle (-1)^{J-M}{-2J\choose -J+M}}
\left (\overline f_1\right )^{J+M}
\left(\overline f_2\right )^{J-M}.
\label{1.15}
\eeq
The binomial factors are defined by the usual continuation of
the Gamma-functions.
This is  similar to  Eq.\ref{1.9}, but here
$M$ is smaller than $-J$ or larger than $J$.  This case of negative
spin is the classical equivalent of the quantum case with negative number
of screening operators. We shall discuss the latter  in section 3,
at the quantum level.

\section{ THE  LIOUVILLE FIELDS WITH POSITIVE
QUANTUM-GROUP SPINS}
\markboth{ 2. Positive spins}{ 2. Positive spins}

\label{2}

In this section, we show how the negative powers of the metric
are reconstructed from the chiral fields whose properties
were  extensively studied recently\cite{G1,G2,G3}.
We are aiming at  quantum versions of Eq.\ref{1.8}.
 As  mentioned above, this discussion was already
summarized in refs.\cite{G4,G5}.
Denote by $C$ the central charge of gravity. The standard screening
charges
$-\alpha_\pm$ of the Liouville theory\cite{GN4,GN6} are such that
$$
\alpha_\pm={1\over 2} \Bigl (\sqrt{ {C-1\over 3}}
\pm \sqrt{ {C-25\over 3}} \Bigr),
$$
\begin{equation}
\alpha_\pm={Q\over 2}\pm \alpha_0, \quad
Q=\sqrt{ {C-1\over 3}}, \quad
\alpha_0={1\over 2} \sqrt{ {C-25\over 3}}
\label{2.1}
\end{equation}
$Q$, and $\alpha_0$ are introduced so that they will agree with the
standard notation, when we couple with matter.
Kac's formula in the present notation is recalled in the
appendix. It  may be written as
\begin{equation}
 \Delta_{Kac} (J,\Jhat;C)=-{1\over 2} \beta(J,\Jhat;C)
\Bigl (\beta(J,\Jhat;C)+Q\Bigr), \quad
\beta(J,\Jhat;C)=J\alpha_-+\Jhat\alpha_+,
\label{2.2}
\end{equation}
where $2J$ and $2\Jhat$ are positive integers.
Thus the most general Liouville field is to  be
written  as $\exp \bigl(-(J\alpha_-+\Jhat\alpha_+)\Phi\bigr)$.
At first we consider the operators
$\exp \bigl(-J\alpha_-\Phi\bigr)$.  Their  chiral components
 have   been extensively studied\cite{G1,G3}.
Their  fusion and braiding are determined by the
Clebsch-Gordan coefficients
and universal R matrix of $U_q(sl(2))$
respectively, with deformation
parameter $h=\pi (\alpha_-)^2/2$, and $q=e^{ih}$.
It may be approriately called
  the Universal Chiral Family (UCF) associated with $U_q(sl(2))$
(There is of
course a parallel discussion for $\exp
\bigl(-\Jhat \alpha_+\Phi\bigr)$,
to which we shall come below).

In refs.\cite{B,G1,G2,G3}, the quantum group properties
are  discussed
 on  the example of the  $x_+$
components---their  characteristic feature is that they
are holomorphic functions of $\tau+i\sigma$. A summary of the
relevent results is given in appendix A for completeness.
 As long as one deals   with functions
of a single  variable, one may describe the whole operator-structure,
without loss of generality, at $\tau=0$,
that is on the unit circle $u=e^{i\sigma}$.
The central charge $C$ and the quantum-group parameter $h$ are
related by
\begin{equation}
C=1+6({h\over\pi}+{\pi\over h}+2).
\label{2.3}
\end{equation}
We work for generic value of $h$, and thus consider irrational
theories. For the rational case,   $q$ is a root of unity, and
the quantum group structure becomes much more complicated. We shall
come back to this point at the end, but we shall see that, although
the formulae we are using at intermediate stages may become meaningless
for rational theories, the final three-point coupling does make
sense. It is thus the correct answer in that case as well.

The UCF is  conveniently described using two different basis of
chiral operators of the type ($1$, $2J+1$), with
$2J$ a positive integer. The first one, called the Bloch-wave
basis,  is made up
with fields denoted    $\psi_m^{(J)}$, $-J\leq m \leq J$,
 that are periodic up to a phase:
\begin{equation}
\psi_m^{(J)}(\sigma+2\pi ) =  e^{2ihm\varpi}
e^{2ihm^2}\,
\psi_m^{(J)}(\sigma),
\label{2.4}
\end{equation}
where the quasi momentum $\varpi$ is an operator such that
\begin{equation}
\psi_m^{(J)}(\sigma)\>\varpi=(\varpi+2m)\>\psi_m^{(J)}(\sigma).
\label{2.5}
\end{equation}
We shall work in a basis where $\varpi$ is diagonal.
$\vert \varpi >$ denotes the corresponding
highest-weight vectors.
The second basis, is made up with operators of the form
$$
\xi_M^{(J)}(\sigma) := \sum_{-J\leq m \leq J}\vert
J,\varpi)_M^m \>
\psi_m^{(J)}(\sigma),  \quad -J\leq M\leq J;
$$
where the $\vert
J,\varpi)_M^m $ are polynomials  of $e^{ih(\varpi
+m)}$ whose expression is recalled in
appendix A. It may be called the quantum group basis, since the
braiding and fusion  relations of the $\xi$ are given by the
universal R matrix and Clebsch-Gordan coefficients of the quantum
group $U_q(sl(2))$ in the standard mathematical form\footnote{The
fusion  property was not really proven  so far, but rather made
very plausible from quantum group invariance\cite{G3}. Its  complete
derivation will be given elsewhere\cite{CGR}. }.

Let us now turn to the $x_-$ components following ref.\cite{G4}
(a similar discussion was independently made
in ref.\cite{B2}). The corresponding
quantities  will be distinguished with a bar: we shall write
$\overline \xi _{M}^{(J)}(x_-)$,
$\overline \psi _{M}^{(J)}(x_-)$,
$\overline \varpi$  and so on.
The  properties of the $x_-$ components are similar to the ones
recalled above and in appendix A for the $x_+$ components, with  a crucial
difference: they are functions of $\tau-i\sigma$, that is
anti-analytic functions of $\tau+i\sigma$. In going from $x_+$ to $x_-$
components, one has to reverse the orientation of the
imaginary axis, which indeed changes  the chirality. The
simplest way to avoid going through the whole argument again,
is to remark that we only need to replace $i$ by $-i$
everywhere in the above formulae, that is, to  use the other root
of $-1$. The whole discussion applies again since it only used
the fact that $i^2+1=0$.
Thus  the quantum-group properties  of $\overline \xi
_{M}^{(J)}(\sigma)$,
$\overline \psi _{M}^{(J)}(\sigma)$ are  the same
as those of $\Bigl ( \xi
_{M}^{(J)}(\sigma)\bigr )^*$,
$\Bigl ( \psi _{M}^{(J)}(\sigma)\bigr )^*$, where the star means
the transposed of the Hermitian conjugate.
 The appropriate definition of
${\overline \xi_M^{(J)}}(\sigma)$ is for instance
\begin{equation}
{\overline \xi_M^{(J)}}(\sigma) := \sum_{-J\leq m \leq J}
\>\bigl(\vert
J,{\overline\varpi})_M^m\bigr)^* \>
{\overline \psi_m^{(J)}}(\sigma),  \quad -J\leq M\leq J;
\label{2.6}
\end{equation}
moreover, the shift propreties  of the $\overline \psi$-fields
are given by
\begin{equation}
{\overline \psi_m^{(J)}}(\sigma)\>\varpi^*
=(\varpi^*+2m)\>{\overline \psi_m^{(J)}}(\sigma).
\label{2.7}
\end{equation}

As is usual in conformally invariant field theory   we assume
that the right- and left-movers commute. Thus
we take the  $\psi$- and $\xi$-fields to commute with  the
$\overline \psi$- and   $\overline \xi$-fields.
(more about this below). The quantum-group structure is
thus a tensor product of the UCF's, which we denote by
$U_q(sl(2))\otimes \overline{U_q(sl(2))}$.

Let us now begin the reconstruction of the Liouville field.
 There are two basic requirements  that determine
$\exp(-J\alpha_- \Phi)$. The first one is  locality, that is,
that it commutes
with any other power of the metric at equal $\tau$. This
condition is required from the consistency of the
quantization according to the scheme recalled in the
introduction, since $\sigma$ is the space variable. In
practice it is also needed so that the correlators of the
Liouville field be single valued around their short-distance
singularities.  The second requirement
 concerns the Hilbert space of states where the physical
operator algebra is realized. The point is that, since
 we took the fields $\xi_{M}^{(J)}$ and $\overline \xi
_{M}^{(J)}$ to commute,  the
quasi momenta $\varpi$ and $\overline \varpi $ of the left- and
right-movers are unrelated, which cannot be true physically, as is
well known.  For reasons that will become clearer later on,
we shall
require that $\exp(-J\alpha_- \Phi)$ leave the subspace
of states with $\varpi=(\overline \varpi)*$ invariant. The
latter condition defines the restriced  Hilbert
${\cal H}_{r}$  where it must be possible to restrict the
operator-algebra  consistently. The restricted Hilbert space is
actually larger than the physical Hilbert space where $\varpi$ is
real, but it is more handy at first, in order to deal with possible
cuts in the $\varpi$-plane.  Remarkably
the two requirements  just stated determine $\exp(-J\alpha_- \Phi)$
almost uniquely. The
appropriate ansatz is:
\begin{equation}
e^{\textstyle -J\alpha_-\Phi(\sigma, \tau )}=
{\tilde c}_J \>a(\varpi)\sum _{M=-J}^J\> (-1)^{J-M}  \>e^{ih(J-M)}\>
\xi_M^{(J)}(x_+)\,
{\overline \xi_{-M}^{(J)}}(x_-) \bigr /a(\varpi)
\label{2.8}
\end{equation}
where ${\tilde c}_J$ is a normalization constant, and $a(\varpi)$ will
be determined below.
Before  starting the derivation,  we
remark that this expression is very natural from the quantum
group viewpoint.  Indeed the $\xi$ fields transform according
to\cite{G1}
\begin{equation}
 J_3\, \xi_{M}^{(J)}=M  \xi_{M}^{(J)},
\quad J_\pm\, \xi_{M}^{(J)}= \sqrt{\lfloor J \mp
M\rfloor
\lfloor J \pm M+1 \rfloor }\, \xi_{M\pm 1}^{(J)}.
\label{2.9}
\end{equation}
so that the  generators $J_\pm$, $J_3$  obey the $U_q(sl(2))$
quantum group algebra
\begin{equation}
\Bigl[J_+,J_-\Bigr]=\lfloor 2J_3 \rfloor, \quad
\Bigl[J_3,J_\pm\Bigr]=\pm J_\pm
\label{2.10}
\end{equation}
Similarly, the transformation of the $\overline \xi$ fields
is given by
\begin{equation}
 {\overline J}_3\,{\overline \xi_{M}^{(J)}}=
M {\overline \xi_{M}^{(J)}},
\quad {\overline J}_\pm\,{\overline \xi_{M}^{(J)}}=
\sqrt{\lfloor J \mp
M\rfloor
\lfloor J \pm M+1 \rfloor }\,{\overline \xi_{M\pm 1}^{(J)}}.
\label{2.11}
\end{equation}
On the other  hand, and as is well known, naive tensor  products
of quantum-group representations do no form representations,
since the algebra is non-linear. It is necessary to use the
co-product. Indeed define
\begin{equation}
 {\cal J}_\pm = J_\pm e^{-ih{\overline J}_3}+e^{ihJ_3}\otimes
{\overline J}_\pm, \quad
 {\cal J}_3 = J_3 + {\overline J}_3,
\label{2.12}
\end{equation}
which does give a representation of Eq.\ref{2.9}. Then one easily
checks that
\begin{equation}
{\cal J}_\pm \exp(-J\alpha_- \Phi)=
{\cal J}_3 \exp(-J\alpha_- \Phi)=0,
\label{2.13}
\end{equation}
so that the quantized Liouville field
is a quantum-group invariant. This
is the quantum version of the classical $sl(2)$ invariance
recalled in the introduction (see Eq.\ref{1.13}). Next,
locality is checked by making use of Eqs.A.26-A.28 which
remain true at equal $\tau\not=0$ since
\begin{equation}
\xi_M^{(J)}(x_+)= e^{\tau L_0}\, \xi_M^{(J)}(\sigma)\,
e^{-\tau L_0}, \quad
{\overline \xi_M^{(J)}}(x_-)=  e^{-\tau{\overline L} _0}
{\overline \xi_M^{(J)}}(\sigma) e^{\tau{\overline L} _0}.
\label{2.14}
\end{equation}
At the present stage, we follow the earlier
conventions \cite{G1,G2,G3} and  define primary fields
on the cylinder $0\leq \sigma
\leq \pi$, $-\infty <\tau <\infty$ in such a way that $\sigma$
and $\tau$ translations are unbroken. This allows in principle
to deal with arbitrary Riemann surface. When we specialize to
the  Riemann sphere, we shall change to the more standard
definition (see below).
Choose  $\pi>\sigma_1>\sigma_2>0 $.
In agreement with the above discussion we have, for instance,
\begin{eqnarray}
{\overline \xi_{M_1}^{(J_1)}}(\sigma_1+i\tau) \,
{\overline \xi_{M_2}^{(J_2)}}(\sigma_2+i\tau)=&\hfill\nonumber \\
\sum_{-J_1\leq N_1\leq J_1;\> -J_2\leq N_2\leq J_2}
\bigl((J_1,J_2)_{M_1\, M_2}^{N_2\, N_1}\bigr)^*&\, {\overline
\xi_{N_2}^{(J_2)}}(\sigma_2+i\tau)) \,
{\overline \xi_{N_1}^{(J_1)}}(\sigma_1+i\tau)),
\label{2.15}
\end{eqnarray}
In checking locality, one encounters the product of two
R   matrices. It is handled by means of the
identities
\begin{equation}
\bigl((J_1,J_2)^{-N_2\,-N_1}_{-M_1\,-M_2}\bigr)^*=
\bigl((J_2,J_1)^{-M_1\,-M_2}_{-N_2\,-N_1}\bigr)^*=
{\overline {(J_2,J_1)}^{M_1\,M_2}_{N_2\,N_1}}
\label{2.16}
\end{equation}
that follow from the explicit expressions Eqs.A.27,
and A.29.
 In this way one deduces the
 equation
\begin{equation}
\sum_{M_1 M_2}
(J_1,J_2)^{P_2\,P_1}_{M_1\,M_2}\>
\bigl((J_1,J_2)^{-N_2\,-N_1}_{-M_1\,-M_2}\bigr)^*
=\delta_{P_1,N_1}\,\delta_{P_2,N_2}
\label{2.17}
\end{equation}
 from the inverse relation Eq.A.30, and the desired
locality relation follows ($a(\varpi)$ does not play any
role so far), that is,
\begin{equation}
e^{\textstyle -J_1\alpha_-\Phi(\sigma_1,\tau)}\,
e^{\textstyle -J_2\alpha_-\Phi(\sigma_2,\tau)}\,=
e^{\textstyle -J_2\alpha_-\Phi(\sigma_2,\tau)}\,
e^{\textstyle -J_1\alpha_-\Phi(\sigma_1,\tau)}
\label{2.18}
\end{equation}
Another requirement is
 that the restricted Hilbert space ${\cal H}_{r}$
be  left invariant. This is verified
  by re-expressing  Eq.\ref{2.8} in
terms of $\psi$ fields. One gets, at first
\begin{eqnarray}
&e^{\textstyle -J\alpha_-\Phi(\sigma, \tau)}=  c_J\,  a(\varpi)
\times&
\nonumber \\
\sum_{M=-J}^J& (-1)^{J-M}\,e^{ih(J-M)}\,\vert J,\,\varpi)_M^m&
\,\bigl(\vert J,\,{\overline \varpi})_{-M}^{p}\bigr)^*
\psi_m^{(J)}(x_+)\,
{\overline \psi_{p}^{(J)}}(x_-) \bigl /  a(\varpi).
\label{2.19}
\end{eqnarray}
Using Eq.A.31,   one writes
\begin{eqnarray}
\sum_{M=-J}^J (-1)^{J-M}\,e^{ih(J-M)}\,\vert J,\,\varpi)_M^m
\,\bigl(\vert J,\,\varpib)_{-M}^{p})^*=\nonumber \\
\sum_{M=-J}^J (-1)^{J-M}\,e^{ih(J-M)}\,\vert J,\,\varpi)_M^m
\,\vert J,\,\varpib^*+2p)_{-M}^{-p}
\label{2.20}
\end{eqnarray}
If $\varpi=\varpib*$, this becomes, according to Eqs.A.18, A.22, and A.23,
\begin{eqnarray}
\sum_{M=-J}^J (-1)^{J-M}\,e^{ih(J-M)}\,\vert J,\,\varpi)_M^m
\,\vert J,\,\varpi+2p)_{-M}^{-p}=
\nonumber \\
(-1)^{J-m}\,
\left (2i \sin (h) e^{ih/2}\right )^{2J}\,
\delta_{m,\,p}\,
{\lambda_m^{(J)}(\varpi)\over \lfloor \varpi+2m\rfloor }.
\label{2.21}
\end{eqnarray}
As a consequence, and when it is restricted to
${\cal H}_{r}$, Eq. \ref{2.8} is equivalent to
\begin{equation}
e^{\textstyle -J\alpha_-\Phi(\sigma, \tau )}=
c_J a(\varpi)\>\sum _{m=-J}^J \, (-1)^{J-m}\,
{\lambda_m^{(J)}(\varpi)\over \lfloor \varpi+2m\rfloor } \>
\psi_m^{(J)}(\sigma)\,
{\overline \psi_{m}^{(J)}}(\sigma) \bigr / a(\varpi),
\label{2.22}
\end{equation}
where $c_J={\tilde c}_J \left (2i \sin (h) e^{ih/2}\right )^{2J}$.
The condition $\varpi*=\overline \varpi$ is indeed left
invariant, in view of   Eqs.\ref{2.5}, and  \ref{2.7}. Finally, and
for reasons that will become  clear in section 4,
 we  choose
$a(\varpi)=1/\sqrt{\lfloor \varpi \rfloor}$. This gives
\begin{equation}
e^{\textstyle -J\alpha_-\Phi(\sigma, \tau )}=
c_J \>\sum _{m=-J}^J \,
{(-1)^{J-m}\, \lambda_m^{(J)}(\varpi)\over
\sqrt{\lfloor \varpi\rfloor} \sqrt {  \lfloor \varpi+2m\rfloor} } \>
\psi_m^{(J)}(x_+)\,
{\overline \psi_{m}^{(J)}}(x_-)
\label{2.23}
\end{equation}
which is the expression we shall use later on. This choice of
$a(\varpi)$ will ensure  the symmetry of the three-point functions, as
we shall see.
At this point a parenthesis is in order. The physical meaning of
the condition $\varpi*=\overline \varpi$ is as follows. When we take
$\varpi$ real at the end,  it will
 mean that left and right movers have equal
momenta. This  will  ensure  that the Liouville field is periodic
in $\sigma$, as it should be, if the
winding number vanishes.  In this connection, let us note that
it is easy to see,
from the explicit expression
Eq.A.19 of $\vert J,\,\varpi)_M^m$,
 that a similar discussion
applies to arbitrary winding number. Then one would have
 $\varpi*=\overline \varpi+r\pi/h$ with
$r$ integer. For simplicity, we only consider the case $r=0$ in
the following.

There remains  to show that the operators  just constructed
 are  closed by fusion.
For this purpose,  the expression Eq.\ref{2.8} is more
handy, since the fusion
properties of the $\xi$ fields are determined by their quantum group
structure\cite{G3,CGR}. One has, on the unit circle,
$$\xi_{M_1}^{(J_1)}(\sigma_1)\,\xi_{M_2}^{(J_2)}(\sigma_2)
= \sum_{J=\vert J_1-J_2\vert}^{J_1+J_2}
\Bigl\{
\bigl(d(\sigma_1-\sigma_2)\bigr)^{\Delta(J)-\Delta(J_1)-\Delta(J_2)}
$$
\begin{equation}
g_{J_1\, J_2}^J
\>\bigl(J_1,M_1;J_2,M_2\vert J_1,J_2;J,M_1+M_2\bigr)\>
\,\Bigl(\xi_{M_1+M_2}^{(J)}(\sigma_1)+\hbox{descendants}\Bigr) \Bigr\},
\label{2.24}
\end{equation}
where $d(\sigma-\sigma')\equiv 1-e^{-i(\sigma-\sigma')}$,
$\bigl(J_1,M_1;J_2,M_2\vert J_1,J_2;J,M_1+M_2\bigr)$ denotes the
Clebsch-Gordan coefficients of $U_q(SL(2))$ (see, e.g. appendix C
of \cite{G3}),
$g_{J_1\, J_2}^J$ are numerical constants, and
$\Delta(J):=-hJ(J+1)/\pi-J$ is the Virasoro-weight of
$\xi_{M}^{(J)}(\sigma)$. The argument given above shows that
the fusion of the $x_-$ components, is given by a similar formula
 the Clebsch-Gordan
 (CG) coefficients are replaced by their complex conjugate.
Closure by fusion will follow from the following identity
\begin{equation}
\bigl(J_1,-M_1;J_2,-M_2\vert J_1,J_2;J,-M_1-M_2\bigr)^*=
\bigl(J_1,M_1;J_2,M_2\vert J_1,J_2;J,M_1+M_2\bigr)
\label{2.25}
\end{equation}
The proof of this relation goes as follows: The explicit expression
of the (CG) coefficients is of the form\cite{G3}
$$
\bigl(J_1,M_1;J_2,M_2 \vert J_1,J_2;J,M\bigr)=
\delta_{M,\,M_1+M_2}\>e^{ih(J_1+J_2-J)(J_1+J_2+J+1)/2}
K(J_1,\,J_2,\,J)\times
$$
$$
e^{ih(M_2 J_1-M_1J_2)}\>
\sum_{\mu=0}^{J_1+J_2-J}\>\Bigl\{{e^{-ih\mu(J+J_1+J_2+1)} \,(-1)^\mu
\over
\lfloor \mu \rfloor\! !\,\lfloor  J_1+J_2-J-\mu \rfloor\! !}\times
$$
\beq
{1\over \lfloor J_1-M_1-\mu \rfloor\! !\,\lfloor J-J_2+M_1+\mu
\rfloor\! !\,
\lfloor  J_2+M_2-\mu \rfloor\! !\,\lfloor J-J_1-M_2+\mu \rfloor\!
!}\Bigr\}.
\label{2.26}
\end{equation}
where $K(J_1,\,J_2,\,J)$ is real. One takes the complex conjugate
of this expression and change $\mu$ into $J_1+J_2-J-\mu $ in the
summation, and the result follows \qed

Closure by
fusion is verified on the unit circle by writing, according to
Eq. \ref{2.24},
$$
e^{\textstyle -J_1\alpha_-\Phi(\sigma_1, 0 )}\>
e^{\textstyle -J_2\alpha_-\Phi(\sigma_2, 0 )}
= a(\varpi)
c_{J_1}c_{J_2}\times
$$
$$\sum_{J=\vert J_1-J_2\vert}^{J_1+J_2}
\sum_{{\overline J}=\vert J_1-J_2\vert}^{J_1+J_2}
\sum _{M_1=-J_1}^{J_1}\> (-1)^{J_1-M_1}  \>e^{ih(J_1-M_1)}\>
\sum _{M_2=-J_2}^{J_2}\> (-1)^{J_2-M_2}  \>e^{ih(J_2-M_2)}\>
$$
$$\Bigl\{
\bigl(d(\sigma_1-\sigma_2)\bigr)^{\Delta(J)-\Delta(J_1)-\Delta(J_2)}
\bigl(d(\sigma_1-\sigma_2)^*\bigr)^
{\Delta({\overline J})-\Delta(J_1)-\Delta(J_2)}
g_{J_1\, J_2}^J (g_{J_1\, J_2}^{{\overline J}})^*
$$
$$\bigl(J_1,M_1;J_2,M_2 \vert J_1,J_2;J,M_1+M_2\bigr)
\>\bigl(J_1,-M_1;J_2,-M_2\vert J_1,J_2;{\overline J},-M_1-M_2)^*\>
$$
\beq
\Bigl(\xi_{M_1+M_2}^{(J)}(\sigma_1)+\hbox{descendants}\Bigr)
\Bigl({\overline \xi}_{-M_1-M_2}^{({\overline J})}(\sigma_1)+
\hbox{descendants}\Bigr)
\Bigr\}
\label{2.27}
\end{equation}
The summation over $M_1$, $M_2$ may be explicitly  carried out for fixed
$M=M_1+M_2$, since it  takes the form
$$
\sum _{M_1+M_2=M}
\bigl(J_1,M_1;J_2,M_2 \vert J_1,J_2;J,M\bigr)
\>\bigl(J_1,-M_1;J_2,-M_2\vert J_1,J_2;{\overline J},-M)^*.
$$
According to Eq.\ref{2.24} one gets
\begin{equation}
\sum _{M_1+M_2=M}
\bigl(J_1,M_1;J_2,M_2 \vert J_1,J_2;J,M\bigr)
\>\bigl(J_1,M_1;J_2,M_2\vert J_1,J_2;{\overline J},M)
=\delta_{J,\,{\overline J}}
\label{2.28}
\end{equation}
where the last equality follws from the orthogonality
of the q CG coeficients\cite{G3}. Thus Eq.\ref{2.27} becomes
$$e^{\textstyle -J_1\alpha_-\Phi(\sigma_1, 0 )}\>
e^{\textstyle -J_2\alpha_-\Phi(\sigma_2, 0 )}
=\sum_{J=\vert J_1-J_2\vert}^{J_1+J_2}
\Bigl\{
\bigl | d(\sigma_1-\sigma_2)\bigr|^{2(\Delta(J)-\Delta(J_1)-\Delta(J_2))}
$$
\beq
{c_{J_1}c_{J_2}\over c_{J}} \bigl | g_{J_1\, J_2}^J \bigr  |^2
\Bigl(e^{\textstyle -J\alpha_-\Phi(\sigma_1, 0 )}+
\hbox{descendants}\Bigr)
\Bigr\}.
\label{2.29}
\eeq
This shows  the closure of  the operator-product expansion at the
level of primaries.

Finally, we come to the most general Liouville field
$\exp \bigl(-(J\alpha_-+\Jhat\alpha_+)\Phi\bigr)$. It is most simply
obtained by fusion from $\exp
\bigl(-J\alpha_-\Phi\bigr)$ and
$\exp \bigl(-\Jhat\alpha_+\Phi\bigr)$. From the quantum-group viewpoint,
the existence of two screening charges comes from the fact that the
relation between $h$ and $C$ (Eq.\ref{2.3}) is quadratic and has two
solutions. Thus there are two quantum-group parameters $h$ and $\hhat$
which are such that
\begin{equation}
C=1+6({h\over\pi}+{\pi\over h}+2)=
1+6({\hhat\over\pi}+{\pi\over\hhat}+2),
\quad \hbox{with} \quad h\hhat=\pi^2,
\label{2.30}
\end{equation}
and $\alpha_+$ is such that $\hhat=\pi (\alpha_+)^2$. The UCF associated
with $\hhat$ is distinguished by hats, and  the hatted counterparts
of Eqs.\ref{2.8} and \ref{2.23} are, respectively,
\begin{equation}
e^{\textstyle -\Jhat \alpha_+\Phi(\sigma, \tau )}=
{\left (2i \sin (\hhat ) e^{i\hhat/2}\right )^{-2\Jhat}\,
{\widetilde{\chat}} _\Jhat\over \sqrt {\varpihat}}
\sum _{\Mhat=-\Jhat}^\Jhat \> (-1)^{\Jhat -\Mhat}  \>e^{ih(\Jhat-\Mhat)}\>
\xihat_\Mhat^{(\Jhat)}(x_+)\,
 \xibhat_{-\Mhat }^{(\Jhat )}(x_-) \sqrt {\varpihat}
\label{2.31}
\end{equation}
\begin{equation}
e^{\textstyle -\Jhat\alpha_+\Phi(\sigma, \tau )}=
\chat_\Jhat \>
\sum _{\mhat=-\Jhat}^\Jhat \, {(-1)^{\Jhat-\mhat}\,
\lambdahat_\mhat^{(\Jhat)}(\varpihat)\over
\sqrt{\lfloorhat \varpihat\rfloorhat}
\sqrt{ \lfloorhat \varpihat+2\mhat\rfloorhat} } \>
\psihat_\mhat^{(\Jhat)}(x_+)\,
 \psibhat_{\mhat}^{(\Jhat)}(x_-)
\label{2.32}
\end{equation}
The momentum $\varpihat$  is defined in complete parallel to $\varpi$,
in order to keep the symmetry between the two quantum group-parameters.
There is really one momentum only and they are proportional
\cite{G1} (see Eq.A.10): $\pi\varpihat=h\varpi$.
 The two UCF's have simple
braiding relations: for $\pi >\sigma_1 >\sigma_2 > 0$ one has\cite{G1}
\begin{equation}
\psi_{m}^{(J)}(\sigma_1)\,
\psihat_{ \mhat }^{(\Jhat) }(\sigma_2)=
e^{-2i\pi J\Jhat }
 \psihat_{ \mhat }^{(\Jhat)  }(\sigma_2)\,
\psi_{m }^{(J) }(\sigma_1),
\label{2.33}
\end{equation}
\begin{equation}
\xi_{m}^{(J)}(\sigma_1)\,
\xihat_{ \mhat }^{(\Jhat) }(\sigma_2)=
e^{-2i\pi J\Jhat }
\>e^{2i\pi (M\Jhat-\Mhat J)}\>
 \xihat_{ \mhat }^{(\Jhat)  }(\sigma_2)\,
\xi_{m }^{(J) }(\sigma_1).
\label{2.34}
\end{equation}
Thus it follows trivially that
\begin{equation}
e^{\textstyle -J_1\alpha_-\Phi(\sigma_1,\tau)}\,
e^{\textstyle -\Jhat_2\alpha_+\Phi(\sigma_2,\tau)}\,=
e^{-4i\pi J_1\Jhat_2 }
e^{\textstyle -\Jhat_2\alpha_+\Phi(\sigma_2,\tau)}\,
e^{\textstyle -J_1\alpha_-\Phi(\sigma_1,\tau)}.
\label{2.35}
\end{equation}
Since $2J_1$ and $2\Jhat_2$ are integers, the factor
$\exp (-4i\pi J_1\Jhat_2 )$ is equal to $1$ and the two set
of fields are mutually local. It is interesting
to note at this point that the condition that $4i\pi J_1\Jhat_2$ be
an integer
  is  similar
 to Dirac's quantization condition for the
product of electric and
magnetic charges.
 There remains to consider the
fusion of the fields with the two values of the screening charges.
The hatted and unhatted fields have simple fusion properties\cite{G1}:
\begin{equation}
\xi_{M_1}^{(J_1)}(\sigma_1) \>
\xihat_{\Mhat_2 }^{(\Jhat_2)}(\sigma_2)\sim
\bigl(d(\sigma_1-\sigma_2)\bigr)^
{\Delta_{Kac}(J_1,\Jhat_2)-\Delta(J_1)-\Deltahat(\Jhat_2)}
e^{i\pi (M_1\Jhat_2-\Mhat_2 J_1)}\>\xi_{ M_1\, \Mhat_2 }^{(J_1\,\Jhat_2) }
(\sigma_1),
\label{2.36}
\end{equation}
where $\xi_{ M_1\, \Mhat_2 }^{(J_1\,\Jhat_2) }$ is the most general
chiral field whose weight is given by Kac's formula.
It immediately follows
from the above formulae that, if we define,
$$e^{\textstyle -(J\alpha_-+\Jhat \alpha_+)\Phi(\sigma, \tau )}=
{{\tilde c}_{J\Jhat}\over \varpi } \times
$$
\beq
\sum _{M, \, \Mhat} \> (-1)^{J-M+\Jhat -\Mhat}
\>e^{ih(J-M)+i\hhat(\Jhat-\Mhat)}\>
\xi_{M\, \Mhat}^{(J\, \Jhat)}(x_+)\,
 \xib_{-M\, -\Mhat }^{(J\, \Jhat )}(x_-) \> \varpi,
\label{2.37}
\eeq
with
$$
{c}_{J\Jhat}=c_J\chat _\Jhat
\left (2i \sin (h) e^{ih/2}\right )^{-2J}
 \left (2i \sin (\hhat ) e^{i\hhat/2}\right )^{-2\Jhat},
$$
we have
$$e^{\textstyle -J\alpha_-\Phi(\sigma_1, \tau )}
\> e^{\textstyle -\Jhat \alpha_+\Phi(\sigma_2, \tau )}\sim
$$
\beq
\bigl\vert d(\sigma_1-\sigma_2)\bigr\vert  ^
{2(\Delta_{Kac}(J,\Jhat)-\Delta(J)-\Deltahat(\Jhat))}\>
e^{\textstyle -(J\alpha_-+\Jhat \alpha_+)\Phi(\sigma_2, \tau )}.
\label{2.38}
\eeq
Closure by fusion and braiding of the most general field
follows from this last equality if one assumes that the order between
fusions  or fusion and braiding is  irrelevent. It may be directly
verified. This is left to the reader.
Finally, another useful expression for the most general Liouville
field may be derived in the Bloch-wave basis.
 The field
$\psi_{ m\, \mhat}^{( J\,\Jhat)}$ is such that\cite{G1}
\begin{equation}
\psi_{m_1}^{(J_1)}(\sigma_1) \>\psihat_{\mhat_2
}^{(\Jhat_2)}(\sigma_2)\sim
\bigl(d(\sigma_1-\sigma_2)\bigr)^
{\Delta_{Kac}(J_1,\Jhat_2)-\Delta(J_1)-\Deltahat(\Jhat_2)}
(-1)^{2(J_1\Jhat_2-m_1\mhat_2)}
\>\psi_{ m_1\, \mhat_2 }^{(J_1\,\Jhat_2)
}
(\sigma_1),
\label{2.39}
\end{equation}
and it follows from Eqs.\ref{2.23}, \ref{2.32}, \ref{2.37} that
\begin{eqnarray}
e^{\textstyle -(J\alpha_-+\Jhat \alpha_+)\Phi(\sigma, \tau )}=
c_{J\Jhat}  \>\sum _{m \, \mhat }&
\nonumber \\
{(-1)^{J-m+\Jhat-\mhat+2(J\Jhat-m\mhat)}
\lambda_m^{(J)}(\varpi)\lambdahat_\mhat^{(\Jhat)}(\varpihat)
\over
\sqrt{ \lfloor \varpi\rfloor}
\sqrt{ \lfloor \varpi+2m+2\mhat \pi/h \rfloor}
\sqrt{ \lfloorhat \varpihat\rfloorhat}
\sqrt{ \lfloorhat \varpihat+2\mhat+2m h/\pi \rfloorhat }}& \>
\psi_{m\, \mhat}^{(J\, \Jhat )}(x_+)\,
 \psib_{m\, \mhat}^{(J\, \Jhat)}(x_-)
\nnn
& \label{2.40}
\end{eqnarray}
The shift of momentum for the general $\psi$ field is given by
\begin{equation}
 \psi_{m\, \mhat}^{(J\, \Jhat )}(\sigma)\> \varpi
=(\varpi+2m+2\mhat\pi/h)\>
\psi_{m\, \mhat}^{(J\, \Jhat )}(\sigma).
\label{2.41}
\end{equation}
Thus the spectrum of eigenvalues of $\varpi$ is of the form
$\varpi^{(0)}+n+\nhat \pi/h$ where $n$ and $\nhat$ are integers, and
$\varpi^{(0)}$ is arbitrary. The $sl(2,C)$ invariant vacuum $\vert
\varpi_0>$, which is such that $L_0 \vert \varpi_0 > =0$ has a
momentum $\varpi_0 =1+\pi/h$ (see Eq.A.11). Physically we should choose
$\varpi^{(0)}=\varpi_0$, and the spectrum of $\varpi$ eigenvalues is
real. At an intermediate stage, however, it is useful to give a
small imaginary part, say positive,  to $\varpi$ and to choose
$\varpi^{(0)}=\varpi_0+i\epsilon$. This allows us  to
handle possible
branch points  in $\varpi$ on the real axis, such
as one sees in Eqs.\ref{2.23}, \ref{2.32}, and
\ref{2.40}.
 The monodromy factor Eq.\ref{2.4}
is such that
\begin{equation}
\exp (2ihm(\varpi_0+n+\nhat \pi/h))
=\exp(2iJ \nhat) \exp (2ihm(\varpi_0+n)).
\label{2.42}
\end{equation}
Thus $\nhat$ describes winding modes (fermionic or bosonic)
with respect to the UCF with parameter $h$.
The general fields $\psi_{m\, \mhat}^{(J\, \Jhat )}(\sigma)$,
or equivalently $\xi_{m\, \mhat}^{(J\, \Jhat )}(\sigma)$,
form a general chiral family (GCF) with a quantum-group structure
$U_q(sl(2))\odot U_\qhat(sl(2))$ where the sign $\odot$
means that the two UCF do not  commute  but satisfy Eqs.\ref{2.33},
\ref{2.34}
instead. The quantum-group structure of the most general
Liouville field is of the type
$U_q(sl(2))\odot U_\qhat(sl(2))\otimes \overline
{U_q(sl(2))\odot U_\qhat(sl(2))}$, where the overlined part
correspond to the minus-components that we took to commute with the
plus-components.

This completes the construction of the Liouville field
on the cylinder. In the coming sections, we shall determine the
three-point functions on the Riemann sphere. Some additional
points are to be made for this purpose,  before closing
the present  section.
The Riemann  sphere,  is described by the complex variable
 $z=\exp (\tau+i\sigma)$. In this connection, let
us recall that a  primary field $A(z, z^*)$ of weight
$\Delta$, $\overline \Delta$  transforms so that
$A(z, z^*) (dz)^\Delta (dz^*)^{\overline \Delta}$ is
invariant\cite{GS}. Given $A(z,z^*)$,  and a conformal
map $z\to Z(z)$, it is thus convenient to define\footnote{
We do not write down the operators
that realize the transformation of the states on which $A$ acts.
This is an abuse of notation made to avoid clumsy formulae.}
\begin{equation}
 A (Z,Z^*) := A(z,z^*) \Bigl ({dz\over dZ}\Bigr )^\Delta
\Bigl ({dz^*\over dZ^*}\Bigr ) ^{\overline \Delta}.
\label{2.43}
\end{equation}
Thus one has
\begin{equation}
A(z,z^*)= e^{-(\tau+i\sigma)\Delta -(\tau-i\sigma) {\overline
\Delta}} A(\sigma, \tau)
\label{2.44}
\end{equation}
 $A(\sigma,\tau)$, which we had been
using before,   was    defined on the
cylinder where the Fourier expansion in  $\sigma$ has
integer coefficients (see, e.g. Eq.A.1). This  rule
gives  the standard
definition for the Laurent expansion of $A(z,z^*)$ as  series in
$z^{n+\Delta}$, and $(z^*)^{m+\overline \Delta}$,  $n$, $m$ integers.
Moreover, $A(z,z^*)$ now transforms covariantly under $z$ and
$z^*$ translations. Thus OPE's take the usual form;   for instance
Eq.\ref{2.29} becomes
\begin{eqnarray}
e^{\textstyle -J_1\alpha_-\Phi(z_1,z^*_1 )}\>
e^{\textstyle -J_2\alpha_-\Phi(z_2,z^*_2 )}
=\sum_{J=\vert J_1-J_2\vert}^{J_1+J_2}
\Bigl\{
\bigl |
z_1-z_2)\bigr|^{2(\Delta(J)-\Delta(J_1)-\Delta(J_2))}\nonumber \\
{c_{J_1}c_{J_2}\over c_{J}} \bigl | g_{J_1\, J_2}^J \bigr
|^2
\Bigl(e^{\textstyle -J\alpha_-\Phi(z_1,z^*_1 )}+
\hbox{descendants}\Bigr)
\Bigr\}.
\label{2.45}
\end{eqnarray}

The last point to discuss is the behaviour of the fields  as
$z$ goes to zero or $\infty$. The situation is not standard,
since we have several fields with the same weights, so that
the connection between primaries and highest-weight states
is not one-to-one.   Concerning the $\psi$  fields
we are going to show that
\begin{eqnarray}
\lim_{z\to 0}   \psi_{m\, \mhat}^{(J\, \Jhat )}(z)
\vert \varpi_0 > \propto \delta_{m+J,\,0}
\> \delta_{\mhat+\Jhat,\,0} \vert \varpi_{J,\,\Jhat} >
\nonumber \\
\lim_{z\to \infty }
<-\varpi_0 \vert
\psi_{m\, \mhat}^{(J\, \Jhat )}(1/ z)
\propto   \delta_{m+J,\,0}\>
\delta_{\mhat+\Jhat,\,0}
< -\varpi_{J,\,\Jhat} \vert,
\label{2.46}
\end{eqnarray}
where
\begin{equation}
 \varpi_{J,\,\Jhat}=\varpi_0+2J+2\Jhat \pi/h.
\label{2.47}
\end{equation}
First, the value of $\varpi_{J,\,\Jhat}$ was to be expected
since Eqs.A.11 and A.33 show that
\begin{equation}
(L_0-\Delta_{Kac}(J,\,\Jhat,\, C))\vert \varpi_{J,\,\Jhat} >=
<-\varpi_{J,\,\Jhat} \vert (L_0-\Delta_{Kac}(J,\,\Jhat,\,
C))=0,
\label{2.48}
\end{equation}
where $\Delta_{Kac}(J,\,\Jhat,\, C)$ is the weight of
$\psi_{m\, \mhat}^{(J\, \Jhat )}$.
Thus  the eigenvalue of $L_0$
is indeed  equal to the weight of the primary field $\psi_{-J\,
-\Jhat}^{(J\, \Jhat )}(z)$. The other primaries
$\psi_{m\, \mhat}^{(J\, \Jhat )}(z)$, $m+J\not= 0$, or
$\mhat+\Jhat\not=0$ have the same weight but their shift are
different. Thus, if the limits were not zero,  states with
wrong eigenvalues of $L_0$ would  come out. This is basically
why Eq.\ref{2.46} gives  zero unless $m+J= 0$,
and $\mhat+\Jhat= 0$.   These relations  may be verified as
follows. By operator-product expansions, the
powers of the field $\psi_{-1/2}^{(1/2)}\equiv \psi_1$
generate the fields $\psi_{-J}^{(J)}$. Since the former is a
simple exponential of the free field $\phi_1$ this is  also
true
for the latter, and one has , according to
 Eq.A.6 with $j=1$,
\begin{eqnarray}
\psi_{-J}^{(J)}(z)\propto N^{(1)}\bigl (e^{2J \sqrt{ h /2\pi}\>
\phi_1}\bigr
)=e^{2J \sqrt{h/2\pi}\, q_0^{(1)}}
z^{2J(\varpi-\varpi_0)h/2\pi}
\nonumber \\
\times \exp\Bigl (2J \sqrt{h/2\pi}
i\sum_{n<0}e^{-in\sigma}p_n^{(1)}/ n\Bigr )
\exp\Bigl (2J \sqrt{h/2\pi}
i\sum_{n>0}e^{-in\sigma}p_n^{(1)}/ n\Bigr)
\label{2.49}
\end{eqnarray}
 This immediately leads to the desired relations
with  $m+J=0$, and $\mhat=\Jhat=0$, using the fact that
for any highest-weight state $\vert  \varpi >$,
\begin{equation}
 p_n^{(j)}\vert \varpi> =
<\varpi \vert p_{-n}^{(j)}=0, \quad n>0, \> j=1,2.
\label{2.50}
\end{equation}
Next, and by the same reasoning,  Eq. (A.6) with
$j=2$ gives
\begin{eqnarray}
\psi_{J}^{(J)}(z)=N^{(2)}\bigl (e^{2J \sqrt{ h /2\pi}\>
\phi_2}\bigr
)=e^{2J \sqrt{h/2\pi}\, q_0^{(2)}}
z^{2J (-\varpi-\varpi_0)h/2\pi}
\nonumber \\
\times \exp\Bigl (2J \sqrt{h/2\pi}
i\sum_{n<0}e^{-in\sigma}p_n^{(2)}/ n\Bigr )
\exp\Bigl (2J \sqrt{h/2\pi}
i\sum_{n>0}e^{-in\sigma}p_n^{(2)}/ n\Bigr )
\label{2.5a}
\end{eqnarray}
Using Eq.\ref{2.50}, with $j=2$, one sees that the limit
Eq.\ref{2.46} vanishes.
This is the
desired result for  $J-m=0$, $\Jhat=\mhat=0$.
A similar discussion obviously deals with the cases
$J=0$, $m=0$, $\Jhat\pm \mhat=0$. Finally,
the result is easily extended to the other values of $m$,
 $\mhat$, and to the general case $J\not=0$, and
$\Jhat\not=0$  by fusion \qed

It follows from  Eqs \ref{2.41} that
\begin{eqnarray}
e^{\textstyle -
(J\alpha_-+\Jhat\alpha_+)\Phi(z,\,z^* )}
\vert \varpi_0 > {{\displaystyle  \sim} \atop z\to 0}&
\nonumber \\
\vert \varpi_{J,\,\Jhat} >
<\varpi_{J,\,\Jhat} \vert &
e^{\textstyle -(J\alpha_-+\Jhat\alpha_+)\Phi(1,\,1 )}
 \vert \varpi_0 >,
\label{2.52}
\end{eqnarray}
\begin{eqnarray}
<-\varpi_0 \vert
e^{\textstyle -(J\alpha_-+\Jhat\alpha_+)\Phi(1/z,\,1/z^* )}
{{\displaystyle \sim} \atop z\to \infty }&
\nonumber \\
 <-\varpi_0 \vert
e^{\textstyle -(J\alpha_-+\Jhat\alpha_+)\Phi(1,\,1 )}&
\vert -\varpi_{J,\,\Jhat} >
<-\varpi_{J,\,\Jhat} \vert
\label{2.53}
\end{eqnarray}
where the matrix elements  on the right-hand side
are given by the  expansion Eq.\ref{2.40} restricted to
$m=-J$,
and $\mhat=-\Jhat$.  One sees that the treatment of the limits
breaks the symmetry between the two free fields $\phi_1$ and
$\phi_2$ which is the basis of the quantum group symmetry.
It is thus  spontaneouly broken by the choice of matrix
elements.
 One may exchange the role of $\phi_1$ and
$\phi_2$, by replacing  $\vert \varpi_0 >$, and
$<-\varpi_0 \vert $ by $\vert -\varpi_0 >$, and
$<\varpi_0 \vert $ respectively.
\section{ THE DRESSING BY GRAVITY}
\markboth{ 3. Dressing by gravity }{ 3. Dressing by gravity }
\label{3}

In this section we study the dressing of conformal models
with central charge $D$ by the Liouville field with central
charge $C$ so that
\begin{equation}
C+D=26.
\label{3.1}
\end{equation}
 We shall be concerned with the case
$D< 1$, where the Liouville theory is in its weakly coupled
regime $C>25$. As is recalled in the appendix A, the existence of the
UCF's  is basically  a consequence of the operator differential
equations A.8, A.9. These  are equivalent to the Virasoro
Ward-identities that describe the decoupling of null vectors. Thus
the UCF's, with appropriate quantum deformation parameters also describe
the matter with $D<1$. We will thus have another copy of the
quantum-group structure recalled above.
It will be distinguished by primes.
Thus we let
\begin{equation}
D=1+6({h'\over\pi}+{\pi\over h'}+2)=
1+6({\hhat'\over\pi}+{\pi\over\hhat'}+2),
\quad \hbox{with} \quad h'\hhat'=\pi^2,
\label{3.2}
\end{equation}
\begin{equation}
h'={\pi \over 12}\Bigl(D-13 -
\sqrt {(D-25)(D-1)}\Bigr),\quad
\hhat'={\pi \over 12}\Bigl(D-13
+\sqrt {(D-25)(D-1)}\Bigr),
\label{3.3}
\end{equation}
and Eq.\ref{3.1} gives
\begin{equation}
h+\hhat'=\hhat+h'=0
\label{3.4}
\end{equation}
Of course we choose the matter and gravity fields to commute.
Using the notation of last section,
the complete quantum group structure is thus of the type:
\begin{eqnarray}
\Bigl\{ \bigl [U_q(sl(2))\odot U_\qhat(sl(2))\bigl ]\otimes \overline
{\bigl [U_q(sl(2))\odot U_\qhat(sl(2))\bigl ]}\Bigr\}\bigotimes
\nonumber \\
\Bigl\{\bigl [ U_{\qhat^{-1}}(sl(2))\odot U_{q^{-1}}(sl(2))\bigl ]
\otimes \overline
{\bigl [U_{\qhat^{-1}}(sl(2))\odot U_{q^{-1}}(sl(2))\bigl ]}\Bigr\},
\label{3.5}
\end{eqnarray}
where the first (second) line  displays the quantum-group structure
of  gravity (matter).
According to the above results, the spectrum of weights
of the gravity and matter are respectively given by
\begin{equation}
 \Delta_{\cal G} (J,\Jhat)={C-1\over 24}-
{ 1 \over 24} \left((J+\Jhat+1) \sqrt{C-1}
-(J-\Jhat) \sqrt{C-25} \right)^2,
\label{3.6}
\end{equation}
\begin{equation}
 \Delta_{\cal M} (J',\Jhat')={D-1\over 24}+
{ 1 \over 24} \left((J'+\Jhat'+1) \sqrt{1-D}
+(J'-\Jhat') \sqrt{25-D} \right)^2.
\label{3.7}
\end{equation}
The connection with Kac's table will be spelled out in section 5.
{}From the standpoint we are taking, the most general matter field is
described by an operator of the form
$\exp \bigl(-(J'\alpha'_-+\Jhat'\alpha'_+)X\bigr)$,
where $X(\sigma,\tau)$ is a local
field that commutes with the Liouville-field
and whose properties are derived from  those of
$\Phi$ by continuation to  central charges smaller than one.
Following our general conventions we let $h'=\pi (\alpha'_-)^2/2$, and
$\hhat'=\pi (\alpha'_+)^2/2$.
The correct screening operator is the field
 $\exp \bigl(-(J\alpha_-+\Jhat\alpha_+)\Phi\bigr)$
with spins $J$ and $\Jhat$ that $\Delta_{\cal G} (J,\Jhat)
+\Delta_{\cal M} (J',\Jhat')=1$.
In this connection, it is an easy consequence of Eq.\ref{3.1} that
\begin{equation}
\Delta_{\cal G} (\Jhat',-J'-1)+\Delta_{\cal M} (J',\Jhat')=1
\label{3.8}
\end{equation}
\begin{equation}
\Delta_{\cal G} (-\Jhat'-1,J')+\Delta_{\cal M} (J',\Jhat')=1
\label{3.9}
\end{equation}
These two  choices  correspond to the existence of two cosmological
terms. Indeed, the unity-operator  of matter ($J'=\Jhat'=0$) is dressed
by operators of the type $\exp (\alpha_+\Phi)$
and $\exp (\alpha_-\Phi)$,
with the choices Eqs.\ref{3.8}, \ref{3.9}, respectively.
As is usual, we choose the
latter as cosmological term so that the spins of gravity and matter
fields will be related by Eq.\ref{3.9}. Thus we shall
be concerned with matrix elements of  operators of the type
\begin{equation}
{\cal V}_{J',\,\Jhat'}(\sigma,\tau)\equiv
 e^{\textstyle ((\Jhat'+1)\alpha_--J' \alpha_+)\Phi(\sigma, \tau )}\>
e^{\textstyle -(J'\alpha'_-+\Jhat' \alpha'_+)
X(\sigma, \tau )}.
\label{3.10}
\end{equation}

Next consider the Hilbert space in which we are working.
The UCF's which
appear in Eq.\ref{3.5} live  in spaces with highest-weight vectors
of the form
\begin{equation}
\vert \varpi > \otimes \vert \varpib  > \otimes
\vert \varpi' > \otimes \vert \varpib'  >.
\label{3.19}
\end{equation}
Indeed, as is hopefully clear from the appendix A (and ref.\cite{G1}),
the two
UCF's  of a product of the type  $U_q(sl(2))\odot U_\qhat(sl(2))$ are
realized in the same Hilbert space. In the restricted  Hilbert space,
one has $\varpi^*=\varpib$, and $\varpi'\, ^*=\varpib'$.
Thus we introduce
sates of the form
\begin{equation}
\vert \varpi,\, \varpi'> \equiv
\vert \varpi > \otimes \vert \varpi^*  > \otimes
\vert \varpi' > \otimes \vert \varpi'\,^*  >.
\label{3.20}
\end{equation}
The next point concerns the physical on-shell states. They should
satisfy the condition
\begin{equation}
(L_0+\overline{L}_0+L_0'+\overline{L}'_0
-2)\vert \varpi,\, \varpi'>=0,
\label{3.21}
\end{equation}
where the notation is self-explanatory. According to Eq.A.11,
this is satisfied if
\begin{equation}
{h\over 4\pi}(1+{\pi\over h})^2-{h\over 4\pi}\varpi^2+
{h'\over 4\pi}(1+{\pi\over h'})^2-{h'\over 4\pi}\varpip^2=1.
\label{3.22}
\end{equation}
 It follows from Eqs.\ref{2.30}, \ref{3.1}, \ref{3.2}, that
\begin{equation}
{h\over 4\pi}(1+{\pi\over h})^2+{h'\over 4\pi}(1+{\pi\over h'})^2
\equiv {C+D-2\over 24}=1.
\label{3.23}
\end{equation}
Moreover, according to Eqs.A.10, and \ref{3.4},
$$
h'
\varpi'^2=\hhat'\varpihatp^2=-h\varpihatp^2
$$
so that, according to Eq.\ref{3.4}, the on-shell condition is
$\varpi^2=\varpihatp^2$. The sign  ambibuity is related with the
 two possibilities of cosmological term.
It will be seen below that
the choice, which is consistent with the above definition of
gravity-dressing is
\begin{equation}
\varpi=-\varpihatp, \quad \hbox{ or, equivalently, } \quad
\varpihat =\varpip.
\label{3.24}
\end{equation}

Our next topic is the precise definition of the operators
${\cal V}_{J',\,\Jhat'}(\sigma,\tau)$.
Concerning matter,  choosing $J'>0$ and $\Jhat'>0$ gives
$\Delta_{\cal M}>0$ and the formulae of last section may be
directly used, obtaining,
$$e^{\textstyle -(J'\alpha'_-+\Jhat' \alpha'_+)X(\sigma, \tau
)}=c_{J',\Jhat'}  \>\sum _{m' \, \mhat' }
(-1)^{J'-m'+\Jhat'-\mhat'+2(J'\Jhat'-m'\mhat')}
\times
$$
$${\lambdap_{m'}^{(J')}(\varpi')\lambdahatp_{\mhat'}^{(\Jhat')}(\varpihat')
\over
\sqrt{ \lceil \varpi'\rceil}
\sqrt{ \lceil \varpi'+2m'+2\mhat' \pi/h\rceil}
\sqrt{ \lceilhat \varpihat'\rceilhat}
\sqrt{ \lceilhat \varpihat'+2\mhat' +2m' h/\pi\rceilhat }} \times
$$
\beq
\chi_{m'\, \mhat'}^{(J'\, \Jhat' )}(x_+)\,
 \overline \chi_{m'\, \mhat'}^{(J'\, \Jhat')}(x_-),
\nonumber \\
\label{3.11}
\eeq
where $\chi$ and $\overline \chi$ are the Bloch-wave operators for
matter, and
where we have let, in general,
\begin{equation}
\lceil y \rceil \equiv {\sin (h'y)\over \sin h'},
\quad \lceilhat y \rceilhat \equiv {\sin (\hhat' y)\over \sin
\hhat'}.
\label{3.12}
\end{equation}
According to Eq.\ref{3.4} one has
\begin{equation}
\lceil x \rceil =\lfloorhat x \rfloorhat, \quad
\lceilhat x \rceilhat =\lfloor  x \rfloor.
\label{3.13}
\end{equation}
Concerning gravity, it has been
 already emphasized\cite{G2,G3}, that the dressing of matter-operators
with positive spins  requires the  use of gravity fields with
negative spins.  For example, the cosmological term
has $J=-1,\Jhat=0$. Thus the gravity-UCF with parameter $h$
 must be extended. This problem
was overcome in \cite{G2,G3} as follows. First, the quantum-group
structure has an obvious symmetry in $J\to -J-1$ with fixed
$M$, so that the
universal R-matrix and Clebsch-Gordan coefficients are left unchanged.
Second, the coefficients $\vert J,\varpi)_M^m$ also
have a natural continuation to negative spin. For postive $J$, one has
\begin{equation}
  \vert -J-1,\varpi)_M^m=
 \vert J,\varpi)_M^m\>(-1)^{J+m}\Bigl /
[(2i\sin
  (h))^{1+2J}\>\lambda_m^{(J)}(\varpi)]
\label{3.14}
\end{equation}
It follows that there exist fields $\xi_M^{(-J-1)}$ with fusion
and braiding similar to $\xi_M^{(J)}$, and fields $\psi_m^{(-J-1)}$
similar to $\psi_m^{(J)}$ so that
\begin{equation}
\xi_M^{(-J-1)}(\sigma)=\sum_{m=-J}^J
{(-1)^{J+m}\>\vert J,\varpi)_M^m \over
((2i\sin
  (h))^{1+2J}\>\lambda_m^{(J)}(\varpi)}\> \psi_m^{(-J-1)}(\sigma)
\label{3.15}
\end{equation}
This motivates us to   introduce a  field of the
form
\begin{equation}
e^{\textstyle (J+1)\alpha_-\widetilde \Phi(\sigma, \tau )}= {1\over
\sqrt \varpi}
{\tilde c}_{-J-1}\sum _{M=-J}^J\> (-1)^{J-M}  \>e^{ih(J-M)}\>
\xi_M^{(-J-1)}(x_+)\,
{\overline \xi_{-M}^{(-J-1)}}(x_-) \>\sqrt \varpi .
\label{3.16}
\end{equation}
We write it as an exponential of  a new field $\widetilde \Phi$,
since it differs from what one expects upon quantization
of the classical expressions Eqs.\ref{1.14}. This field satisfies
all the necessary requirements.
First it  is local since the exchange properties
of the fields $\xi_{M}^{(-J-1)}$ and
${\overline \xi_{M}^{(-J-1)}}$ are the same as those of the fields
$\xi_{M}^{(J)}$, and
${\overline \xi_{M}^{(J)}}$,  respectively. Second, it may also be
re-expressed in the Bloch-wave basis, making use of Eq.\ref{3.14}. One
finds, taking Eq.\ref{2.20} into account,
\begin{equation}
e^{\textstyle (J+1)\alpha_-\widetilde \Phi(\sigma, \tau )}=
c_J \>\sum _{m=-J}^J \,
{1\over \lambda_m^{(J)}(\varpi)
\sqrt{\lfloor \varpi\rfloor} \sqrt{  \lfloor \varpi+2m\rfloor} } \>
\psi_m^{(-J-1)}(x_+)\,
{\overline \psi_{m}^{(-J-1)}}(x_-),
\label{3.17}
\end{equation}
which shows that this field leaves the condition $\varpi=\varpib$
invariant.  Remarkably, we will find that the range of $m$ which
appears in this last equation is precisely what is needed to compute
matrix elements between physical states satisfying condition
\ref{3.24}. This fully motivates the introduction of the field
$\widetilde \Phi$.  By fusion, one finally arrives at the expression of the
dressing-operator
\begin{eqnarray}
&e^{\textstyle ((J+1)\alpha_-\widetilde \Phi(\sigma, \tau
)-\Jhat \alpha_+\Phi(\sigma, \tau
))}=c_{-J-1\, \Jhat}
  \>{\displaystyle \sum _{m \, \mhat }}  \,
\nonumber \\
&{\lambdahat_\mhat^{(\Jhat)}(\varpihat)
\over  \lambda_m^{(J)}(\varpi)\,
\sqrt{ \lfloor \varpi\rfloor}
\sqrt{ \lfloor \varpi+2m+2\mhat \pi/h\rfloor}
\sqrt{ \lfloorhat \varpihat\rfloorhat}
\sqrt{ \lfloorhat \varpihat+2\mhat+2mh/\pi \rfloorhat }} \>
\psi_{m\, \mhat}^{(-J-1\, \Jhat )}(x_+)\,
 \psib_{m\, \mhat}^{(-J-1\, \Jhat)}(x_-).
\nonumber \\
\label{3.18}
\end{eqnarray}

Next we want   to compute the matrix element of the operator
${\cal V}_{J',\,\Jhat'}(\sigma,\tau)$, given by Eq.\ref{3.10},
between highest-weight states. For this
purpose, we shall make use of the expressions in terms of $\psi$-fields
Eqs.\ref{3.11}, \ref{3.18}. The matrix element of a  typical operator
$\psi_{m\> \mhat} ^{(J\> \Jhat)}$ between highest weight states
was fully  determined
in \cite{G1} for positive $J$. In Appendix E of this reference,
quantities denoted $C^{\mu,\nu;\>\muhat,\nuhat}$ and
$D^{\mu,\nu;\>\muhat,\nuhat}(\varpi)$ were
defined
 so that
\begin{equation}
\psi^{(J\>\Jhat)}_{m\>\mhat}(\sigma)\equiv
C^{\mu,\nu;\>\muhat,\nuhat}\>
D^{\mu,\nu;\>\muhat,\nuhat}(\varpi)\>\>
 V^{\mu,\nu;\>\muhat,\nuhat}(\sigma),
\label{3.25}
\end{equation}
where
\begin{equation}
2J=\mu+\nu,\quad  2m= \nu-\mu,\quad  2\Jhat=\muhat+\nuhat,
 \quad 2\mhat= \nuhat-\muhat.
\label{3.26}
\end{equation}
 $V^{\mu,\nu;\>\muhat,\nuhat}(\sigma)$
is an operator whose normalization does not depend upon its
indices: its only non-vanishing matrix element is
\begin{equation}
< \varpi  \vert \,V^{\mu,\nu;\>\muhat,\nuhat}(\sigma)
\vert \varpi-\mu+\nu+(-\muhat+\nuhat)\pi/h >=
e^{i\sigma (\varpi^2-(\varpi-\mu+\nu+(-\muhat+\nuhat)\pi/h)^2)h/4\pi}\>.
\label{3.27}
\end{equation}
The fact that the shift of $\varpi$ can take only one value follows
from  Eq.A.12.  According to Eqs.A.16 and E.3 of
\cite{G1}, $C^{\mu,\nu;\>\muhat,\nuhat}$ is given
by
\begin{equation}
C^{\mu,\nu;\>\muhat,\nuhat}=
C^{\mu,\nu}\Chat^{\muhat,\nuhat}\>\>
{\prod_{r=1}^{\mu+\nu}\prod_{\rhat=1}^{\muhat+\nuhat}
\bigl(r\sqrt{h/ \pi}+\rhat\sqrt{\pi/h}\bigr) \over
\Bigl [\prod_{s=1}^{\mu}\prod_{\shat=1}^{\muhat}
\bigl(s\sqrt{h/ \pi}+\shat\sqrt{\pi/h}\bigr)\,
\prod_{t=1}^\nu\prod_{\that=1}^\nuhat
\bigl(t\sqrt{h/ \pi}+\that\sqrt{\pi/h}\bigr)\Bigr ]},
\label{3.28}
\end{equation}
\begin{equation}
 C^{\mu, \nu}=\prod_{r=1}^{\mu+\nu}
 \Gamma\bigl(1+rh/ \pi\bigr)\Bigl /\Bigl [
\prod_{s=1}^{\mu}\Gamma\bigl(1+sh/\pi\bigr)\,
\prod_{t=1}^\nu \Gamma\bigl(1+th/\pi\bigr)\Bigr ],
\label{3.29}
\end{equation}
with similar formulae for $\Chat^{\muhat,\nuhat}$. The $\varpi$
dependent part $D^{\mu,\nu;\>\muhat,\nuhat}(\varpi)$ is
given\footnote{ up to a few misprints,}
  by EqsA.26a,b and E.4 of \cite{G1}:
\begin{equation}
D^{\mu,\nu;\>\muhat,\nuhat}(\varpi) =
D^{\mu,\nu}(\varpi)\>\Dhat^{\muhat,\nuhat}(\varpi) \>
M^{\mu,\nu;\>\muhat,\nuhat}(\varpi)
\label{3.30}
\end{equation}
\begin{equation}
 D^{\mu, \nu} (\varpi)=
\prod_{t=1}^{\mu-\nu} {\sqrt{\Gamma\bigl(-(\varpi-t+1)h/\pi
\bigr)}\over
\sqrt{\Gamma((\varpi-t) h/ \pi)}}
\prod_{r=1}^\mu \Gamma\bigl((\varpi-r) h/\pi )\bigr)
\prod_{s=1}^\nu \Gamma\bigl((-\varpi-s) h/\pi )\bigr),
\label{3.31}
\end{equation}
if $\mu>\nu$, and
\begin{equation}
 D^{\mu, \nu} (\varpi)=
\prod_{t=1}^{\nu-\mu} {\sqrt{\Gamma\bigl((\varpi+t-1)h/\pi\bigr)}
\over
\sqrt{ \Gamma((-\varpi-t) h/ \pi)}}
\prod_{r=1}^\mu \Gamma\bigl((\varpi-r) h/\pi )\bigr)
\prod_{s=1}^\nu \Gamma\bigl((-\varpi-s) h/\pi )\bigr),
\label{3.32}
\end{equation}
if $\nu>\mu$. Moreover, if
for instance, $\mu>\nu$, $\muhat>\nuhat$,
$$ M^{\mu,\nu;\>\muhat,\nuhat}(\varpi)=
$$
\beq
{{\prod_{t=1}^{\mu-\nu}\prod_{\that=1}^{\muhat-\nuhat}
\sqrt{-(\varpi-t+1)\sqrt{h/ \pi}+(\that-1)\sqrt{\pi/h}}
\sqrt{(\varpi-t)\sqrt{h/ \pi}-\that\sqrt{\pi/h}}}\over
{
\prod_{r=1}^{\mu}\prod_{\rhat=1}^{\muhat}
\bigl((\varpi-r)\sqrt{h/ \pi}-\rhat\sqrt{\pi/h}\bigr)
\>\prod_{s=1}^{\mu}\prod_{\shat=1}^{\nuhat}
\bigl(-(\varpi+s)\sqrt{h/ \pi}-\shat\sqrt{\pi/h}\bigr)}}.
\label{3.33}
\eeq
Eqs.\ref{3.31} and \ref{3.32} correspond to a situation which one
encounters repeatedly, namely, one goes from one to the other
by continuing a product of the form $\prod_{j=1}^N f_j$ to negative
values of $N$. The answer, which is well known in mathematics is
obtained by writing (this trick was already used in \cite{G3})
$$
\prod_{j=1}^N f_j=\prod_{j=1}^A f_j\>\Bigl/\prod_{k=N+1}^{A} f_k,
$$
where $A$ is an integer larger than $N$. This leads
to the rule
\begin{equation}
\prod_{j=1}^N f_j\Longrightarrow \prod_{j=N+1}^0 {1\over f_j}
\equiv \prod_{j=1}^{-N} {1\over f_{1-j}}, \quad
\hbox{ for}\quad N\leq -1.
\label{3.34}
\end{equation}
Eqs.\ref{3.31},  and \ref{3.32} are an example of this rule. It also
applies, as one may verify from the derivation of \cite{G1},
in order  to
define $M^{\mu,\nu;\>\muhat,\nuhat}$ when $\mu-\nu$, and/or
$\muhat-\nuhat$  are not positive. {\bf In the following,  we shall
freely write products from $1$ to negative numbers, with the
understanding that they are defined from the rule Eq.\ref{3.34}.}

At this point, a  technical point must be straighten out: The
expressions we wrote, such as Eqs.\ref{3.31}-\ref{3.33} involve square
roots of functions of $\varpi$,
and we have to take account of the  branch-points.
They are for
real eigenvalues  of $\varpi$. This is why we gave to $\varpi$
an imaginary part to be removed at the end of the calculation.
The derivation of   Eqs.\ref{3.31}-\ref{3.33}
of \cite{G1} did not explicitly specify the definitions
of the square roots. In order to do so, as simply as possible,
we remark that it is based on the solution of recurrence relations
which may be reduced to the basic identity
\begin{equation}
\sqrt{\Gamma(z+1)} =
\sqrt z \sqrt{\Gamma(z)},
\label{3.35}
\end{equation}
 for an arbitrary complex variable $z$.
By convention, we choose the usual definition $\sqrt z =
\sqrt \rho \exp (i\theta/2)$, where $zz^*=\rho^2$, and where $
-\pi <\theta < \pi$ is the argument of $z$. In this way,
Eq.\ref{3.35} allows us to uniquely define
$\sqrt{\Gamma(z)}$ in the whole complex plane with cuts between
$-2N-2$ and $-2N-1$, $N$ positive integer. Then the function
$\sqrt {\sin (\pi z)}$ is defined so that
\begin{equation}
\sqrt{\Gamma(z)} \sqrt{\Gamma(1-z)} =
{\sqrt \pi \over \sqrt {\sin (\pi z)}}.
\label{3.36}
\end{equation}
The imaginary part of  $\varpi$
is unchanged by
the $\psi$-fields which shift $\varpi$
by real amounts. It
removes the sign ambiguity for square roots of negative
real numbers that may appear.
This precaution is necessary at the present intermediate stage
only, however, since  we shall see that all cuts  disappear
from the final answer. On the other hand,
 the appearence of cuts spoils the general
argument, recalled in the appendix (Eqs.A.14, and A.32), that
formally shows that the exponentials  of the Liouville field are
hermitian operators. This point already made briefly in ref.\cite{G1}
is related to the objections  to the construction of the Liouville
field which were  raised in ref.\cite{BP}.

Formulae Eqs.\ref{3.25} -- \ref{3.33}
directly apply to the matter-field
Eq.\ref{3.11}, after making the suitable replacements by primed
quantities. Since $h'<0$, one may be worried at
first sight by the factors $\sqrt {h/\pi}$
that occur in Eqs.\ref{3.28} and
\ref{3.33}. An easy computation shows that the product
$C^{\mu,\nu;\>\muhat,\nuhat}
M^{\mu,\nu;\>\muhat,\nuhat}(\varpi)$ is a rational function of
$h$. There is thus no  problem in replacing $h$ by $h'$.
 It is  convenient to use expressions similar to the above
 that are   symmetric between unhatted and
hatted quantities.  We adopt the following definitions
\begin{equation}
\Cp^{\mu',\nu';\>\muhat',\nuhat'}=
{\Cp^{\mu',\nu'}\Chatp^{\muhat',\nuhat'}
 \prod_{r=1}^{\mu'+\nu'}\prod_{\rhat=1}^{\muhat'+\nuhat'}
\bigl(r\sqrt{\pi/h}-\rhat\sqrt{h/ \pi}\bigr) \over
\prod_{s=1}^{\mu'}\prod_{\shat=1}^{\muhat'}
\bigl(s\sqrt{ \pi/h }-\shat\sqrt{h/\pi}\bigr)\,
\prod_{t=1}^{\nu'}\prod_{\that=1}^{\nuhat'}
\bigl(t\sqrt{ \pi/h}-\that\sqrt{h/\pi}\bigr)},
\label{3.37}
\end{equation}
 $$M^{\mu',\nu';\>\muhat',\nuhat'}(\varpi')=
$$
\beq
{{\prod_{t=1}^{\mu'-\nu'}\prod_{\that=1}^{\muhat'-\nuhat'}
\sqrt{-(\varpi'-t+1)\sqrt{\pi/h}-(\that-1)\sqrt{h/\pi}}
\sqrt{(\varpi'-t)\sqrt{\pi/h}+\that\sqrt{h/\pi}}}\over
{
\prod_{r=1}^{\mu'}\prod_{\rhat=1}^{\muhat'}
\bigl((\varpi'-r)\sqrt{\pi/h}+\rhat\sqrt{h/\pi}\bigr)
\>\prod_{s=1}^{\mu'}\prod_{\shat=1}^{\nuhat'}
\bigl(-(\varpi'+s)\sqrt{\pi/h}+\shat\sqrt{h/\pi}\bigr)}},
\label{3.38}
\eeq
whose  product is the correct continuation of Eqs.\ref{3.28}, and
\ref{3.33}.

 For gravity we
need to continue to negative $J$. This was the aim of theorem (5.1)
of \cite{G3}. The basic point is that the derivation of
Eqs. \ref{3.28}-\ref{3.33} carried out in \cite{G1} is based on the
hypergeometric differential
equation which is obeyed by the two-point functions $<\varpi_2\vert
 \psi_{\pm
1/2}^{(1/2)}(\sigma_1) \psi^{(J\>\Jhat)}_{m\>\mhat}(\sigma_2) \vert
\varpi_1 > $, and $<\varpi_2\vert
 \psihat_{\pm
1/2}^{(1/2)}(\sigma_1) \psi^{(J\>\Jhat)}_{m\>\mhat}(\sigma_2) \vert
\varpi_1 > $.
This differential equation is a consequence of Eqs.A.8 and A.9 valid
for any sign of $J$ and $\Jhat$. Thus the coefficients
$C^{\mu,\nu;\>\muhat,\nuhat}$, and
$D^{\mu,\nu;\>\muhat,\nuhat}(\varpi)$ obey the same recursion
relations (solved in appendix E of \cite{G1}) after continuation in
$J$ or $\Jhat$. With an appropriate choice of their overall
normalisations, they are thus given by the continuation of
Eqs.\ref{3.28} -- \ref{3.33}
according to the rule Eq.\ref{3.34}.

Our aim is to compute the three-point function of the form
\begin{equation}
\bigl <
\prod_{\ell=1}^3 {\cal V}_{J'_\ell,\,\Jhat'_\ell}(z_\ell,\,z^*_\ell)
 \bigr > =
{\cal C}_{1,2,3}\Bigr /
\Bigl (\prod_{k,l} \vert z_k-z_l \vert^2 \Bigr  ).
\label{3.39}
\end{equation}
where ${\cal C}_{1,2,3}$ is the coupling  constant.
  Since the fields
${\cal V}_{J'_l,\,\Jhat'_l}(z_l,\,z^*_l)$ are local, their order
should be  irrelevant on the left-hand side. We nevertheless fix it
temporarily in order to perform the calculation.
As usual, we shall take
 $z_3\to \infty$,
 $z_2=1$, $z_1=0$, and consider the matrix element between
the $sl(2,C)$-invariant states $<-\varpi_0, -\varpi'_0\vert$ and
$\vert  \varpi_0, \varpi'_0 >$.  Then  ${\cal C}_{1,2,3}$ will be
computed  from the
matrix elements between highest-weight sates that come
out following
 a reasoning similar  to the one that
led to Eqs.\ref{2.52} and \ref{2.53}. The part of gravity with negative
spin introduces a new subtlety in this connection, since the
$sl(2,C)$-invariant states do not satisfy the physical condition
Eq.\ref{3.24} (this is obvious since they are annihilated by $L_0$,
$L'_0$, $\overline L_0$, and $\overline L'_0$).  The
derivation of Eqs.\ref{2.52} and \ref{2.53}  showed that the  limits
only involve the terms that are simple exponentials of the field
$\phi_1$, that is,  those for which $m$, and  $\mhat$ are equal to
minus the unhatted and the hatted spins, respectively.
Those terms are absent from $\exp[ (J+1)\alpha_-\widetilde \Phi]$,
as defined by Eq.\ref{3.16}, and thus we conclude that
\begin{equation}
\lim_{z\to 0} e^{\textstyle (J+1)\alpha_-\widetilde \Phi(z,\,z^* )}
\vert \varpi_0 >=
\lim_{z\to \infty }
<-\varpi_0 \vert
e^{\textstyle (J+1)\alpha_-\widetilde \Phi(1/z,\,1/z^* )}=0.
\label{3.40}
\end{equation}
We need to define another field to create the highest-weight
states of the spectrum from the $sl(2,C)$-invariant left and
right vacua. Denoted by
$e^{\textstyle (J+1)\alpha_- \Phi(z,\,z^* )}$, it is  such that
the limit is finite:
\begin{eqnarray}
e^{\textstyle
(J+1) \alpha_- \Phi(z,\,z^* )}
\vert \varpi_0 >{{\textstyle  \sim} \atop z\to 0}
\nonumber \\
\vert
\varpi_{-J-1,\,0} >
<\varpi_{-J-1,\,0} \vert
e^{\textstyle (J+1)\alpha_- \Phi(1,\,1 )}
 \vert \varpi_0 >,
\label{3.41}
\end{eqnarray}
\begin{eqnarray}
<-\varpi_0 \vert
e^{\textstyle (J+1)\alpha_- \Phi(1/z,\,1/z^* )}
{{\textstyle \sim} \atop z\to \infty }
\nonumber \\
 <-\varpi_0 \vert
e^{\textstyle (J+1)\alpha_- \Phi(1,\,1 )}
\vert -\varpi_{J,\,0} >
<-\varpi_{J,\,0} \vert.
\label{3.42}
\end{eqnarray}
We define   matrix elements  between the highest-weight
states that appear  in the last equations, as  given
by the continuation of Eq.\ref{2.40}
to $J=\pm m$. It is appropriate to
 write these fields  as exponentials of the original Liouville field
$\Phi$ since they  correspond to the quantum
versions  of the classical expression Eq.\ref{1.15}.
Finally  the coupling constants are defined  by
$${\cal C}_{1,2,3}=
<-\varpi_0, -\varpi'_0\vert  {\widetilde {\cal V}}_
{J'_3,\,\Jhat'_3} \vert  -\varpi_3, -\varpi'_3 >\times
$$
\beq
 <-\varpi_3, -\varpi'_3\vert    {\cal V}_
{J'_2,\,\Jhat'_2} \vert  \varpi_1, \varpi'_1 >
<\varpi_1, \varpi'_1\vert  {\widetilde {\cal V}}_
{J'_1,\,\Jhat'_1} \vert  \varpi_0, \varpi'_0 >,
\label{3.43}
\eeq
where  the  notation is such that the complete symmetry
between the three operators will be manifest at the end.
The operators denoted with a tilde are given  by
\begin{equation}
{\widetilde {\cal V}}_{J',\,\Jhat'}(\sigma,\tau)\equiv
 e^{\textstyle (\Jhat'+1)\alpha_-{ \widetilde\Phi}(\sigma, \tau )
 -J' \alpha_+\Phi(\sigma, \tau )
}\>
e^{\textstyle -(J'\alpha'_-+\Jhat' \alpha'_+)
X(\sigma, \tau )}.
\label{3.44}
\end{equation}
By definition they are such that Eqs.\ref{3.41} and \ref{3.42}
do not give zero. By construction,  the corresponding matrix
elements are given by the same formulae as for   ${\widetilde {\cal
V}}_{J',\,\Jhat'}$, after continuation to the appropriate
quantum numbers (more about this below).

It is convenient to introduce $\mu$  and $\nu$  variables similar
to Eq.\ref{3.26} for each leg. According to Eqs.\ref{2.53}, and
\ref{2.47},
the parameters relevant for the first and third operators satisfy,
for $l=1$, and $3$,
\begin{eqnarray}
2J_l=\mu_l,\quad 2\Jhat_l=\muhat_l,\quad&
2J'_l=\mu'_l,\quad
2\Jhat'_l=\muhat'_l
\nonumber \\
\mu_l=-\muhat'_l-2,\quad \muhat_l=\mu_l',\quad&
\nu_l=\nuhat'_l= \muhat_l=\mu_l'=0,
\nonumber \\
\varpi_l=\varpi_0+\mu_l+\muhat_l {\pi\over h},\quad
\varpi'_l=\varpi'_0+\mu'_l-\muhat'_l {h\over \pi} ,
\quad & \varpi_l=-\varpihat'_l,\quad
\varpihat_l=\varpi'_l
\label{3.45}
\end{eqnarray}
For ${\cal
V}_{J'_2,\,\Jhat'_2}$ the only possible values  of $m_2$, $\mhat_2$,
 $m'_2$, and $\mhat_2$  are such that
\begin{equation}
\varpi_1+\varpi_3=2m_2+2\mhat_2\pi/h,\quad
\varpi'_1+\varpi'_3=2m'_2-2\mhat'_2h/\pi
\label{3.46}
\end{equation}
Since $h$ is not rational,  this leads to
\beqa
\nu_2=1+J_1+J_2+J_3, \quad &\nuhat_2=1+\Jhat_1+\Jhat_2+\Jhat_3,
\nnn
\quad \nu'_2=1+J'_1+J'_2+J'_3, &
\nuhat'_2=1+\Jhat'_1+\Jhat'_2+\Jhat'_3.
\label{3.47}
\eeqa
 $\nu_2, \cdots \nuhat'_2$ are equal to
the number of screening operators in the Coulomb
gas language.
According to Eq.\ref{3.10}, one also has
\begin{equation}
J_l=-\Jhat'_l-1,\quad \Jhat_l=J'_l,
\quad l=1,\, 2,\, 3.
\label{3.48}
\end{equation}
Combined with  the previous relations, this gives
\begin{equation}
\mu_2=-\muhat_2'-1,\quad \muhat_2=\mu'_2, \quad
\nu_2=-\nuhat_2'-1,\quad \nuhat_2=\nu'_2
\label{3.49}
\end{equation}
It is easy to see that, as a consequence of this last equation,
the choice $-\Jhat'_2 \leq \mhat'_2 \leq \Jhat'_2$  gives
$J_2+1 \leq m_2 \leq -J_2+1$ which is precisely the range
of values which appear in the definition Eq.\ref{3.17}. Thus the
field $\widehat \Phi$ obtained from the symmetry between
spins  $J$ and $-J-1$ is precisely the one needed for the
proper dressing by gravity, when  dealing with  the mid point.
\section{ HIGHEST-WEIGHT MATRIX ELEMENTS}
\markboth{4. Matrix elements} {4. Matrix elements}
\label{4}

Our aim in this section is to compute the three  matrix elements
that appear in Eq.\ref{3.43}. We shall temporarily
use a simplified notation.
These matrix elements  are
of the form
\beq
<\varpi,\, \varpip \vert \,
{\cal V}_{J',\,\Jhat'}
\vert \varpi_i,\, \varpi_i >, \> \hbox{or}\>
<\varpi,\, \varpip \vert \,
{\widetilde {\cal V}} _{J',\,\Jhat'}
\vert \varpi_i,\, \varpi_i >
\label{4.1}
\eeq
According to Eq.\ref{3.48}, the spins of the gravity part is given by
\begin{equation}
\Jhat=J',\quad \hbox{and}\quad  J=-\Jhat'-1
\label{4.2}
\eeq
The calculation cannot be done for all three terms simultaneously,
since they are of different types. It is convenient
to compute the middle one first, since the other two will
be obtained after  simple modifications.

\subsection{ Matrix elements between on-shell states}

The characteristic feature of the matrix element
$<-\varpi_3, -\varpi'_3\vert  {\cal
V}_{J'_2,\,\Jhat'_2} \vert  \varpi_1, \varpi'_1 >$
is that it is computed between states that both satisfy
condition Eq.\ref{3.24}, and are thus on-shell states satisfying
Eq.\ref{3.21}. Thus our first  problem is to compute Eq.\ref{4.1}, with
\beq
\varpi=-\varpihatp, \quad \varpi_i=-\varpihatp_i,
\quad \varpihat =\varpip, \quad \varpihat_i =\varpip_i.
\label{4.3}
\eeq
In this subsection,  we write ${\cal
V}_{J',\,\Jhat'}$  instead of  ${\cal
V}_{J'_2,\,\Jhat'_)2}$, and    Eq.\ref{3.48}.
becomes
\beq
\mu=-\muhat'-1,\quad \muhat=\mu', \quad
\nu=-\nuhat'-1,\quad \nuhat=\nu'.
\label{4.4}
\eeq
In performing the calculation, one makes use of
Eqs. \ref{3.28}-\ref{3.33}, and \ref{3.37}, \ref{3.38}. It is
convenient to put together the corresponding
gravity and matter pieces, since striking  simplifications take
place. For instance, one has:
$$D^{\mu, \nu} (\varpi)\Dhatp^{\muhat', \nuhat'} (\varpihat')=
D^{\mu, \nu} (\varpi)\Dhatp^{-\mu-1, -\nu-1} (-\varpi)=
$$
$$\prod_{t=1}^{\mu-\nu} {\sqrt{\Gamma\bigl(-(\varpi-t+1)h/\pi
\bigr)}\over
\sqrt{\Gamma((\varpi-t) h/ \pi)}}
\prod_{t=1}^{-\mu+\nu} {\sqrt{\Gamma\bigl(-(\varpihat'-t+1)\hhat'/\pi
\bigr)}\over
\sqrt{\Gamma((\varpihat'-t) \hhat'/ \pi)}}
$$
$$\prod_{r=1}^\mu \Gamma\bigl((\varpi-r) h/\pi )\bigr)
\prod_{r=1}^{-\mu-1} \Gamma\bigl((\varpihat'-r) \hhat'/\pi )\bigr)
$$
\beq
\prod_{s=1}^\nu \Gamma\bigl((-\varpi-s) h/\pi )\bigr)
\prod_{s=1}^{-\nu-1} \Gamma\bigl((-\varpihat'-s) \hhat'/\pi )\bigr).
\label{4.5}
\eeq
The terms are combined using rule Eq.\ref{3.34}. For instance, the first
line gives, according to Eqs.\ref{3.35}, and \ref{3.36},
$$\prod_{t=1}^{\mu-\nu} {\sqrt{\Gamma\bigl(-(\varpi-t+1)h/\pi
\bigr)}\over
\sqrt{\Gamma((\varpi-t) h/ \pi)}}
{\sqrt{\Gamma\bigl((\varpi-t+1)h/\pi
\bigr)}\over
\sqrt{\Gamma(-(\varpi-t) h/ \pi)}} =
$$
\beq
{\sqrt{-(\varpi-\mu+\nu)} \sqrt{\sin (\varpi-\mu+\nu)}
\over \sqrt{-\varpi} \sqrt{\sin \varpi}},
\label{4.6}
\eeq
and the first term of the second line becomes
\beq
\prod_{r=1}^\mu \Gamma\bigl((\varpi-r) h/\pi )\bigr)
\prod_{r=1}^{\mu+1}{1\over  \Gamma\bigl((\varpi-r+1) h/\pi )\bigr)}=
{1\over \Gamma\bigl(\varpi h/\pi )},
\label{4.7}
\eeq
and so on. Altogether, one finds
\beq
D^{\mu, \nu} (\varpi)\Dhatp^{\muhat', \nuhat'} (\varpihat')=
{h\sin h \over \pi^2} \sqrt{-\varpi} \sqrt{-(\varpi-\mu+\nu)}
 \sqrt{\lfloor \varpi\rfloor } \sqrt{\lfloor \varpi-\mu+\nu\rfloor}
\label{4.8}
\eeq
\beqa
\Dhat^{\muhat, \nuhat} (\varpihat)
\Dp^{\mu', \nu'} (\varpi')=
\left ({-\pi^2\over \hhat \sin \hhat}\right )^{\muhat+\nuhat}
\nnn
{\sqrt{-(\varpihat-\muhat+\nuhat)}\over \sqrt{-\varpihat}}
{\sqrt{\lfloorhat \varpihat-\muhat+\nuhat\rfloorhat}
\sqrt{\lfloorhat \varpihat\rfloorhat} \over
\lfloorhat \varpihat-\muhat\rfloorhat_{\muhat+\nuhat+1}}
\prod_{\rhat=1}^\muhat{1\over ( \varpihat-\rhat)}
\prod_{\shat=1}^\nuhat {1\over (\varpihat+\shat)}
\label{4.9}
\eeqa
\beq
M^{\mu, \nu} (\varpi)\Mhatp^{\muhat', \nuhat'} (\varpihat')
=e^{i\varphi_1}  \left ( {\pi\over h}\right )^{(\muhat+\nuhat)/2}
{ \sqrt{\varpi} \sqrt{\varpi_i}
\prod_{\rhat=1}^\muhat (\varpihat-\rhat)
\prod_{\shat=1}^\nuhat (\varpihat+\shat)
\over
\sqrt{\varpi-(\muhat-\nuhat)\pi/h}
\sqrt{\varpi-\mu+\nu}}
\label{4.10}
\eeq
\beq
e^{i\varphi_1}= \prod _{t=1}^{\mu-\nu}
\prod _{\that=1}^{\muhat-\nuhat}
{\sqrt{-(\varpi-t+1)\sqrt{h/ \pi}+(\that-1)\sqrt{\pi/h}}
\over \sqrt{(\varpi-t+1)\sqrt{h/ \pi}-(\that-1)\sqrt{\pi/h}}}
{\sqrt{(\varpi-t)\sqrt{h/ \pi}-\that\sqrt{\pi/h}}\over
\sqrt{-(\varpi-t)\sqrt{h/ \pi}+\that\sqrt{\pi/h}}}
\label{4.11}
\eeq
\beqa
D^{\mu,\nu;\>\muhat,\nuhat}(\varpi)
\Dp^{\mu',\nu';\>\muhat',\nuhat'}(\varpi')=
\nnn
e^{i\varphi_2} {h\sin h \over \pi^2}
\Bigl ({\sqrt{ \pi h} \over  \sin \hhat}\Bigr )^{\muhat+\nuhat}
{\sqrt{\varpi} \sqrt{\varpi_i}\
\sqrt{\lfloor
\varpi\rfloor} \sqrt{\lfloor \varpi_i\rfloor}
\sqrt{\lfloorhat
\varpihat\rfloorhat} \sqrt{\lfloorhat \varpihat_i\rfloorhat}
\over  \lfloorhat \varpihat-\muhat\rfloorhat_{\muhat+\nuhat+1}}
\label{4.12}
\eeqa
\beqa
e^{i\varphi_2}=e^{i\varphi_1}
{\sqrt{-(\varpihat-\muhat+\nuhat)}\over
\sqrt{\varpihat-\muhat+\nuhat}}
{\sqrt{-(\varpi-\mu+\nu)}\over
\sqrt{\varpi-\mu+\nu}} \times
\nnn
{ \sqrt{\lfloorhat \varpihat-\muhat+\nuhat\rfloorhat}\over
\sqrt{\lfloorhat \varpihat-\muhat+\nuhat
-(\mu-\nu)h/\pi\rfloorhat}}
{\sqrt{\lfloor \varpi-\mu+\nu\rfloor}\over
\sqrt{\lfloor \varpi-\mu+\nu-(\muhat-\nuhat)\pi/h\rfloor}}
\label{4.13}
\eeqa
\beq
C^{\mu,\nu;\>\muhat,\nuhat}
\Cp^{\mu',\nu';\>\muhat',\nuhat'}
=\left ( {h\over \pi} \right ) ^{(\muhat+\nuhat)/2}
{\lfloorhat \muhat \rfloorhat \! ! \lfloorhat \nuhat \rfloorhat \! !
\over \lfloorhat \muhat+\nuhat \rfloorhat \! !}
{1\over \Gamma \Bigl [1+\muhat+\nuhat+(1+\mu+\nu)h/\pi\Bigr ]}
\label{4.14}
\eeq
Making use of Eqs.\ref{3.25} and \ref{3.26}, this leads to
$$<\varpi\vert \psi_{m,\,\mhat}^{(J,\,\Jhat)}(1)\vert \varpi_i>
<\varpi'\vert \psip_{m',\,\mhat'}^{(J',\,\Jhat')}(1)\vert \varpi'_i>
={h\sin h \over \pi^2}
\left ({h\over  \sin \hhat}\right )^{\muhat+\nuhat}
e^{i\varphi_2}
{\lfloorhat \muhat \rfloorhat \! ! \lfloorhat \nuhat \rfloorhat
\! !
\over \lfloorhat \muhat+\nuhat \rfloorhat \! !} \times
$$
\beq
{\sqrt{\varpi} \sqrt{\varpi_i}
\sqrt{\lfloor
\varpi\rfloor} \sqrt{\lfloor \varpi_i\rfloor}
\sqrt{\lfloorhat
\varpihat\rfloorhat} \sqrt{\lfloorhat \varpihat_i\rfloorhat}
\over  \lfloorhat \varpihat-\muhat\rfloorhat_{\muhat+\nuhat+1}
\Gamma \Bigl [1+\muhat+\nuhat+(1+\mu+\nu)h/\pi\Bigr ]}
\label{4.15}
\eeq

Finally, one combines Eqs.\ref{3.10}, \ref{3.11},
\ref{3.18}, with the last
equation,  and writes
\beqa
<\varpi,\, \varpip \vert \,
{\cal V}_{J',\,\Jhat'}
\vert \varpi_i,\, \varpi_i >= c_{J,\,\Jhat}\cp_{J',\,\Jhat'}
\times
\nnn
{\lambdahat_{\mhat}^{(\Jhat)}(\varpihat)\>
\lambdap_{m'}^{(J')}(\varpi')\>
\lambdahatp_{\mhat'}^{(\Jhat')}(\varpihat') \>
\over \lambda_{m}^{(-J-1)}(\varpi) \>
\sqrt {\lfloor \varpi   \rfloor }
\sqrt {\lfloor  \varpi_i \rfloor }
\sqrt {\lfloorhat \varpihat \rfloorhat }
\sqrt {\lfloorhat \varpihat_i \rfloorhat }
\sqrt {\lceil \varpi' \rceil }
\sqrt {\lceil \varpi'_i \rceil }
\sqrt {\lceilhat \varpihat' \rceilhat }
\sqrt {\lceilhat \varpihat'_i \rceilhat }}\times
\nnn
<\varpi,\, \varpip \vert \,\Bigl \{
\psi_{m,\,\mhat}^{(-\Jhat'-1,\,-J')}(1)\>
\psib_{m,\,\mhat}^{(-\Jhat'-1,\,-J')}(1)\>
\psip_{m',\,\mhat'}^{(J',\,\Jhat')}(1)\>
\psib_{m',\,\mhat'}^{(J',\,\Jhat')}(1)\>
\Bigr \} \vert \varpi_i,\, \varpip_i >
\label{4.16}
\eeqa
It follows from the discussions carried out earlier that
\beq
m=-\mu+\nu,\quad \mhat =-\muhat+\nuhat, \quad
m'=-\mu'+\nu',\quad \mhat' =-\muhat'+\nuhat',
\label{4.17}
\eeq
so that one has, according to Eq.\ref{4.2},
\beq
m=-\mhat',\quad \mhat=m'.
\label{4.18}
\eeq
This gives
\beq
\lambdap_{m'}^{(J')}(\varpi')=
\lambdahat_{\mhat}^{(\Jhat)}(\varpihat), \quad
\lambda_{m}^{(-J-1)}(\varpi)=
(-1)^{2\Jhat'+1} \lambdahatp_{\mhat'}^{(\Jhat')}(\varpihat'),
\label{4.19}
\eeq
and Eq.\ref{4.16} becomes
\beqa
<\varpi,\, \varpip \vert \,
{\cal V}_{J',\,\Jhat'}(\sigma,\tau)
\vert \varpi_i,\, \varpi_i >=
{c_{J,\,\Jhat}\>\cp_{J',\,\Jhat'}
\bigl (\lambdahat_{\mhat}^{(\Jhat)}(\varpihat)\bigr)^2\>
\over \lfloor \varpi   \rfloor
 \lfloor  \varpi_i \rfloor
\lfloorhat \varpihat \rfloorhat
\lfloorhat \varpihat_i \rfloorhat } (-1)^{2\Jhat'+1}
\times
\nnn
<\varpi,\, \varpip \vert \,\Bigl \{
\psi_{m,\,\mhat}^{(-\Jhat'-1,\,-J')}(1)\>
\psib_{m,\,\mhat}^{(-\Jhat'-1,\,-J')}(1)\>
\psip_{m',\,\mhat'}^{(J',\,\Jhat')}(1)\>
\psib_{m',\,\mhat'}^{(J',\,\Jhat')}(1)\>
\Bigr \} \vert \varpi_i,\, \varpip_i >,
\label{4.20}
\eeqa
where
\beq
e^{i\varphi(\varpi,\varpi_i)}=
{\sqrt {\lfloor \varpi   \rfloor } \over
\sqrt {-\lfloor \varpi   \rfloor } }
{\sqrt {\lfloor \varpi_i   \rfloor } \over
\sqrt {-\lfloor \varpi_i   \rfloor } }.
\label{4.21}
\eeq
Substituting Eq.\ref{4.14} and the analogous expression with
bars, one arrives at a very simple expression
\beq
<\varpi,\, \varpip \vert \,
{\cal V}_{J',\,\Jhat'}(\sigma,\tau)
\vert \varpi_i,\, \varpi_i >=
B_{J',\,\Jhat'}\>
e^{i\varphi(\varpi,\varpi_i)}
{\varpi \varpi_i \over
\left (\Gamma\Bigl[1+\muhat+\nuhat+(1+\mu+\nu)h/\pi\Bigr ]\right)
^2},
\label{4.22}
\eeq
\beq
B_{J',\,\Jhat'}=  (-1)^{2J'+1}
c_{J,\,\Jhat}\>\cp_{J',\,\Jhat'}
\left ({h\sin h \over \pi^2}\right )^2
\left ({h\over  \sin \hhat}\right )^{\Jhat}
\label{4.23}
\eeq
One sees that the final result is completely symmetric between $\varpi$
and $\varpi_i$. This justifies a posteriori our choice of
$a(\varpi)$ in defining the Liouville field (Eqs.\ref{2.22},
\ref{2.23})

\subsection{ Matrix element with the right vacuum}

There are two  differences between the matrix element $<\varpi_1,
\varpi'_1\vert   {\cal
V}_{J'_1,\,\Jhat'_1} \vert  \varpi_0, \varpi'_0 >$ and the one
we just computed. First,  the state $
\vert  \varpi_0, \varpi'_0 >$ has weights $\Delta(\varpi_0)=
\Delta(\varpi_0')=0$ so that it does not satisfy the
on-shell condition Eq.\ref{3.21}. Second,  the $\nu$'s vanish.
The problem is now to compute Eq.\ref{4.1}, with,
\beq
\varpi=-\varpihatp, \quad \varpi_i=1+\pi/h,
\quad \varpihat =\varpip, \quad \varpihatp_i =1-h/\pi,
\label{4.24}
\eeq
and, according to Eq.\ref{3.48},
\beq
\mu=-\muhat'-2,\quad \muhat=\mu', \quad
\nu=\nuhat'=0,\quad \nuhat=\nu'=0.
\label{4.25}
\eeq
One again computes  $<\varpi\vert
\psi_{m,\,\mhat}^{(J,\,\Jhat)}(1)\vert \varpi_i>
<\varpi'\vert \psip_{m',\,\mhat'}^{(J',\,\Jhat')}(1)\vert
\varpi'_i>$. It is possible  to separate  terms that are
identical to the previous calculation, with the $\nu$'s set
to zero, from  new factors which must be computed. The result
are most simply presented in the same way. Eqs.\ref{4.8}, \ref{4.9},
and \ref{4.10} are respectively replaced by
$$D^{\mu, 0} (\varpi)\Dhatp^{\muhat', 0}
(\varpihat')=\Bigl [\hbox {expression (4.8)}\Bigr ]
\times
$$
\beq
{\sqrt{\Gamma\bigl((\varpi-\mu)h/\pi
\bigr)}\over
\sqrt{\Gamma\bigl(-(\varpi-\mu-1) h/ \pi\bigr)}}
{\sqrt{\Gamma\bigl((\varpi-\mu-1)h/\pi
\bigr)}\over
\sqrt{\Gamma\bigl (-(\varpi-\mu-2) h/ \pi\bigr )}}
{\Gamma\bigl(-\varpi h/\pi
\bigr)\over
\Gamma\bigl ((\varpi-\mu-1) h/ \pi\bigr)}
\label{4.26}
\eeq
\beq
\Dhat^{\muhat, 0} (\varpihat)
\Dp^{\mu', 0} (\varpi')=
\Bigl [\hbox {expression (4.9)}\Bigr ]
\label{4.27}
\eeq
$$M^{\mu, 0} (\varpi)\Mhatp^{\muhat', 0} (\varpihat')
= \Bigl [\hbox {expression (4.10)}\Bigr ]
\times
$$
$${\sqrt{\Gamma\bigl(-\varpihat+(\mu+1)h/\pi
\bigr)}\over
\sqrt{\Gamma\bigl (-\varpihat+\muhat+(\mu+1) h/ \pi\bigr)}}
{\sqrt{\Gamma\bigl(-\varpihat+(\mu+2)h/\pi
\bigr)}\over
\sqrt{\Gamma\bigl (-\varpihat+\muhat+(\mu+2) h/ \pi\bigr )}}
{\sqrt{\Gamma\bigl(\varpihat-\muhat-\mu h/\pi
\bigr)}\over
\sqrt{\Gamma\bigl (\varpihat-\mu h/ \pi\bigr )}}
\times
$$
\beq
{\sqrt{\Gamma\bigl(\varpihat-\muhat-(\mu+1)h/\pi
\bigr)}\over
\sqrt{\Gamma\bigl (\varpihat-(\mu+1) h/ \pi\bigr )}}
{\Gamma\bigl(\varpihat-(\mu+1) h/\pi
\bigr)\over
\Gamma\bigl (\varpihat-\muhat-(\mu+1) h/ \pi\bigr )}
\left ({h\over \pi}\right )^{\muhat/2},
\label{4.28}
\eeq
and, combining the last three relations
$$D^{\mu,0;\>\muhat,0}(\varpi)
\Dp^{\mu',0;\>\muhat',0}(\varpi')=
\Bigl [\hbox {expression (4.12)}\Bigr ]
\left ({h\over \pi}\right )^{\muhat/2}
\times
$$
\beq
{\Gamma\bigl(-\varpihat \bigr)
\sqrt{\Gamma\bigl(\varpihat-\muhat-\mu h/\pi
\bigr)}\over
\sqrt{\Gamma\bigl (-\varpihat+\muhat+(\mu+1) h/ \pi\bigr )}
\sqrt{\Gamma\bigl (-\varpihat+\muhat+(\mu+2) h/ \pi\bigr )}
\Gamma\bigl (\varpihat-\muhat-(\mu+1) h/ \pi\bigr )}.
\label{4.29}
\eeq
Using the fact that $\varpihat-\muhat-\mu h/ \pi=\varpihat_0$,
 one finds
\beqa
D^{\mu,0;\>\muhat,0}(\varpi)
\Dp^{\mu',0;\>\muhat',0}(\varpi')=
\Bigl [\hbox {expression (4.12)}\Bigr ]
\times
\nnn
\left ({h\over \pi}\right )^{\muhat/2}
\Gamma\bigl(-\varpihat \bigr)
{\sqrt{\Gamma\bigl(\varpihat_i
\bigr)}\over
\sqrt{\Gamma\bigl (-\varpihat_i+h/\pi)}
\sqrt{\Gamma\bigl (\varpihat_i-h/\pi\bigr )}
\sqrt{\Gamma\bigl (-\varpihat_i+2h/\pi\bigr )}}.
\label{4.30}
\eeqa
 On the other hand, Eqs.\ref{3.28}, and \ref{3.29}  immediately
show that
\beq
C^{\mu,0;\>\muhat,0}
\Cp^{\mu',0;\>\muhat',0}=1,
\label{4.31}
\eeq
and this replaces Eq.\ref{4.14} of the previous calculation.
The matrix element of the $\psi$ fields is now given by
\beqa
<\varpi\vert \psi_{m,\,\mhat}^{(J,\,\Jhat)}(1)\vert \varpi_i>
&<\varpi'\vert \psip_{m',\,\mhat'}^{(J',\,\Jhat')}(1)\vert
\varpi'_i>
={h\sin h \over \pi^2}
\left ({h\over  \sin \hhat}\right )^{\muhat}
e^{i\varphi_2}
{\Gamma\bigl(-\varpihat \bigr)
\over  \lfloorhat \varpihat-\muhat\rfloorhat_{\muhat+1}}
\times
\nnn
&{\sqrt{\Gamma\bigl(\varpihat_i
\bigr)}
\sqrt{\varpi} \sqrt{\varpi_i}
\sqrt{\lfloor
\varpi\rfloor} \sqrt{\lfloor \varpi_i\rfloor}
\sqrt{\lfloorhat
\varpihat\rfloorhat} \sqrt{\lfloorhat \varpihat_i\rfloorhat}
\over
\sqrt{\Gamma\bigl (-\varpihat_i+h/\pi)}
\sqrt{\Gamma\bigl (\varpihat_i-h/\pi\bigr )}
\sqrt{\Gamma\bigl (-\varpihat_i+2h/\pi\bigr )}}
\label{4.32}
\eeqa
Assuming that  the field $\widetilde \Phi$ is given by the same
expression as $\Phi$, one again makes use of Eq.\ref{4.16}
(there is a change of normalisation which will appear soon).
The choice of $m$ variables is now
\beq
m=-J,\quad, \mhat=-\Jhat,\quad
m'=-J',\quad, \mhat'=-\Jhat'.
\label{4.33}
\eeq
According to Eq.\ref{4.24}, and this last relation,
one has
\beq
\lambdap_{-J'}^{(J')}(\varpi')=
\lambdahat_{-\Jhat}^{(\Jhat)}(\varpihat), \quad
\lambda_{-J}^{(-J-1)}(\varpi)=
{\lfloor \varpi_i-1\rfloor \over
\lfloor \varpi \rfloor ^2 }
(-1)^{2\Jhat'+1} \lambdahatp_{-\Jhat'}^{(\Jhat')}(\varpihat').
\label{4.34}
\eeq
The derivation of the second relation needs  some explanation.
One first writes, according to  Eqs.A.23, and  \ref{4.33},
\beq
\lambda_{-J}^{(-J-1)}(\varpi)=\lambda_{\Jhat'+1}^{(\Jhat')}(\varpi)
={\lfloor 2\Jhat' \rfloor \! ! \over
\lfloor 2\Jhat'+1 \rfloor \! !
\lfloor -1 \rfloor \! !}
\lfloor  \varpi+1 \rfloor_{2\Jhat'+1}
\label{4.35}
\eeq
Since in general $\lfloor N \rfloor \! !$ is defined as
the solution of $\lfloor N+1 \rfloor \! ! =
 \lfloor N+1\rfloor \lfloor N \rfloor \! !$, it follows that
$\lfloor -1 \rfloor \! !=0$, and we remove it once for
all by changing the overall normalisation.
On the other hand,
$$ \lambdahatp_{-\Jhat'}^{(\Jhat')}(\varpihat')=
\lceilhat \varpihat'-2\Jhat'\rceilhat _{2\Jhat'+1}
=\lfloor -\varpi-2\Jhat'\rfloor _{2\Jhat'+1},
$$
and, comparing the last two equalities,
$$\lambda_{-J}^{(-J-1)}(\varpi)=
\lambdahatp_{-\Jhat'}^{(\Jhat')}(\varpihat') \>
(-1)^{2\Jhat'+1}
{\lfloor \varpi+2\Jhat'+1\rfloor \over
\lfloor 2\Jhat'+1\rfloor
\lfloor \varpi\rfloor}.
$$
It is easy to see that
$$\lfloor \varpi+2\Jhat'+1\rfloor=
\lfloor \varpi_i-1\rfloor (-1)^{2\Jhat},\quad
\lfloor 2\Jhat'+1\rfloor=\lfloor \varpi\rfloor
(-1)^{2\Jhat},
$$
and the second relation Eq.\ref{4.34} follows. From this point,
the calculations proceeds
in the same way as above, and one gets
$$<\varpi,\, \varpip \vert \,
{\cal V}_{J',\,\Jhat'}
\vert \varpi_i,\, \varpi_i >=
$$
\beq
{B_{J',\,\Jhat'}\>
e^{i\varphi(\varpi,\varpi_i)}
\Gamma\bigl(\varpihat_i
\bigr) \over \lfloor \varpi_i-1\rfloor
\Gamma\bigl (-\varpihat_i+h/\pi \bigr )
\Gamma\bigl (\varpihat_i-h/\pi \bigr )
\Gamma\bigl (-\varpihat_i+2h/\pi \bigr )}
\varpi \varpi_i \left [\Gamma\bigl(-\varpihat \bigr)
 \lfloor \varpi \rfloor \right ]^2.
\label{4.36}
\eeq
One writes
$$\lfloor \varpi_i-1\rfloor
\Gamma\bigl (-\varpihat_i+h/\pi \bigr )=
{\sin \left ( \pi \bigl (\varpihat_i -h/\pi \bigr )\right )
\Gamma\bigl (-\varpihat_i+h/\pi \bigr )
\over \sin h },
$$
and it follows from the relations $\Gamma(z) \Gamma(1-z)=\pi /
\sin (\pi z )$ and  $\varpihat_i =1+h/\pi$ that the
above expression is simply equal to $\pi /\sin h$. Using
moreover the identity
$$\Gamma\bigl(-\varpihat \bigr)
 \lfloor \varpi \rfloor =
{-\pi \over \varpihat \Gamma (\varpihat) \sin h},
$$
one finally derives the desired expression
\beq
<\varpi,\, \varpip \vert \,
{\widetilde {\cal V}}_{J',\,\Jhat'}
\vert \varpi_i,\, \varpi_i >=
B_{J',\,\Jhat'}\>
 e^{i\varphi(\varpi,\varpi_i)}
\Gamma\bigl(1+h/\pi
\bigr)
\Gamma\bigl (2-h/\pi \bigr )
 {\varpi_i\over \varpi} {1\over \left [\Gamma\bigl(\varpihat \bigr)
   \right ]^2}.
\label{4.37}
\eeq
This result is perfectly finite, but this is not true at
intermediate stages of the calculation: in Eq.\ref{4.36},
$ \lfloor \varpi_i-1\rfloor$ is actually equal to
zero for $\varpi_i=1+\pi /h$, while $\Gamma\bigl
(-\varpihat_i+h/\pi \bigr )=\Gamma(-1)$ diverges. One  may
bypass the ambiguity by
 first giving a small imaginary part
to $\varpi_0$, as needed for the treatment of the possible
cuts.

\subsection{ Matrix element with the left vacuum}

This last case is relevent for the computation of $<-\varpi_0,
-\varpi'_0\vert {\cal
V}_{J'_3,\,\Jhat'_3} \vert  -\varpi_3, -\varpi'_3 >$ in
Eq.\ref{3.43}. The calculation is performed in the same way as for
the case of subsection 4.2. One now has, instead of Eq.\ref{4.24},
\beq
\varpi=-1-\pi/h,\quad \varpi_i=-\varpi'_i, \quad
\varpihat=-1-h/\pi,\quad \varpihat_i=\varpi.
\label{4.38}
\eeq
Since the $\nu$ variables are zero, Eq. \ref{4.25} remains valid.
Computing again the modifications, with respect to the case
4.1, one has to take account of the fact that the relationship
between $\varpi$ and $\varpi'$ is modified to
\beq
\varpihat'=-\varpi-2,\quad
\varpi'=\varpihat+2h/\pi.
\label{4.39}
\eeq
Eqs.\ref{4.8}, \ref{4.9},
and Eqs.\ref{4.10} are respectively replaced by
\beqa
D^{\mu, 0} (\varpi)\Dhatp^{\muhat', 0}
(\varpihat')=\Bigl [\hbox {expression (4.8)}\Bigr ]
{\sqrt{\Gamma\bigl((\varpi+2)h/\pi
\bigr)}\over
\sqrt{\Gamma\bigl (-(\varpi+1) h/ \pi \bigr )}}
\times
\nnn
\sqrt{\Gamma\bigl((\varpi+1)h/\pi
\bigr)}
\sqrt{\Gamma\bigl (-\varpi h/ \pi\bigr )}
{\Gamma\bigl((\varpi-\mu) h/\pi
\bigr)
\over\Gamma\bigl ((\varpi+1) h/ \pi\bigr)
\Gamma\bigl ((\varpi+2) h/ \pi\bigr)}
\label{4.40}
\eeqa
\beqa
\Dhat^{\muhat, 0} (\varpihat)
\Dp^{\mu', 0} (\varpi')=
\Bigl [\hbox {expression (4.9)}\Bigr ]
\times \nnn
\prod_{t=1}^{\muhat}
{\sqrt{\Gamma\bigl((\varpihat-t+1)\pi/h+2
\bigr)}\over
\sqrt{\Gamma((\varpihat-t+1)\pi/h\bigr )}}
{\sqrt{\Gamma\bigl(-(\varpihat-t)\pi/h-2
\bigr)}\over
\sqrt{\Gamma(-(\varpihat-t)\pi/h\bigr )}}
\label{4.41}
\eeqa
\beqa
M^{\mu, 0} (\varpi)\Mhatp^{\muhat', 0} (\varpihat')
= \Bigl [\hbox {expression (4.10)}\Bigr ]
\times
\nnn
 \left ({h\over \pi}\right )^{\muhat/2}
e^{i\varphi_4} \prod_{t=1}^{\muhat}
{\sqrt{\Gamma\bigl(\varpi-(t-1)\pi/h
\bigr)}\over
\sqrt{\Gamma\bigl (\varpi-(t-1)\pi/h+2\bigr)}}
{\sqrt{\Gamma\bigl(-\varpi+t\pi/h-2
\bigr)}\over
\sqrt{\Gamma\bigl (-\varpi+t\pi/h\bigr)}}
\times
\nnn
{\Gamma\bigl((\varpi-\mu) h/\pi-\muhat
\bigr)
\Gamma\bigl((\varpi+2) h/\pi
\bigr)
\Gamma\bigl((\varpi+1) h/\pi
\bigr)
\over
\Gamma\bigl (\varpi-\mu) h/ \pi\bigr )
\Gamma\bigl (\varpi+2) h/ \pi-\muhat \bigr )
\Gamma\bigl (\varpi+1) h/ \pi-\muhat \bigr )},
\label{4.42}
\eeqa
where
\beqa
e^{i\varphi_4} =
\prod_{t=1}^{\muhat} \Bigl \{
{\sqrt{\varpi+1+(t-1)\pi/h
}\over
\sqrt{-\varpi-1-(t-1)\pi/h}}
\times \nnn
{\sqrt{\varpi+(t-1)\pi/h
}\over
\sqrt{-\varpi-(t-1)\pi/h}}
{\sqrt{\varpi+2-t\pi/h
}\over
\sqrt{-\varpi-2+t\pi/h}}
{\sqrt{\varpi+1-t\pi/h
}\over
\sqrt{-\varpi-1+t\pi/h}}\Bigr\}.
\label{4.43}
\eeqa
Combining Eqs.\ref{4.40}, \ref{4.41}, and \ref{4.42},  one gets,  after a
calculation similar to the one that lead to Eq.\ref{4.30}
\beqa
 D^{\mu,0;\>\muhat,0}(\varpi)
\Dp^{\mu',0;\>\muhat',0}(\varpi')=
\Bigl [\hbox {expression (4.12)}\Bigr ]
\times
\nnn
\left ({h\over \pi}\right )^{\muhat/2}\>
e^{i\varphi_4}
\Gamma\bigl(\varpihat_i \bigr)
{\sqrt{\Gamma\bigl(-\varpi h/\pi
\bigr)}\over
\sqrt{\Gamma\bigl ((\varpi+1)h/\pi)}
\sqrt{\Gamma\bigl (-(\varpi+1)h/\pi)}
\sqrt{\Gamma\bigl ((\varpi+2)h/\pi\bigr )}}.
\label{4.44}
\eeqa
 The end of the calculations is exactly as in 4.2. Using
Eq.\ref{4.39}, one verifies  that
Eqs.\ref{4.34}  are replaced by
\beq
\lambdap_{-J'}^{(J')}(\varpi')=
\lambdahat_{-\Jhat}^{(\Jhat)}(\varpihat), \quad
\lambda_{-J}^{(-J-1)}(\varpi)=
{\lfloor \varpi+1\rfloor \over
\lfloor \varpi_i \rfloor ^2 }
(-1)^{2\Jhat'+1} \lambdahatp_{-\Jhat'}^{(\Jhat')}(\varpihat').
\label{4.45}
\eeq
The final result is
\beq
<\varpi,\, \varpip \vert \,
{\widetilde {\cal V}}_{J',\,\Jhat'}
\vert \varpi_i,\, \varpi_i >=
B_{J',\,\Jhat'}\>
 e^{i\varphi(\varpi,\varpi_i)}
\Gamma\bigl(1+h/\pi
\bigr)
\Gamma\bigl (2-h/\pi \bigr )
{\varpi \over \varpi_i}
{1\over \left [\Gamma\bigl(-\varpihat_i \bigr)\right ]^2}
\label{4.46}
\eeq
\section{ THE THREE-POINT COUPLING}
\markboth{3.  Three-point}{3. Three-point}
\label{5}
The three-point function is finally derived by substituting
Eq.\ref{4.22} (with $\varpi=-\varpi_3$, $\varpi_i=\varpi_3$,
and $\mu+\nu=J_2$, $\cdots$, $\muhat'+\nuhat'=\Jhat'_2$),
Eq.\ref{4.37} (with $\varpi=\varpi_1$, $\varpi_i=\varpi_0$,
and $J=J_2$, $\cdots$, $\Jhat,=\Jhat'_2$),
  and Eq.\ref{4.46} (with $\varpi=-\varpi_0$, $\varpi_i=-\varpi_3$,
and $J=J_3$, $\cdots$, $\Jhat,=\Jhat'_3$),
 into Eq.\ref{3.43}. One gets
$${\cal C}_{1,2,3}=-
e^{i\varphi(-\varpi_0,-\varpi_3)}
e^{i\varphi(-\varpi_3,\varpi_1)}
e^{i\varphi(\varpi_1,\varpi_0)}
\left (\Gamma\bigl(1+h/\pi
\bigr)
\Gamma\bigl (2-h/\pi \bigr ) \right )^2
\times$$
\beq
\prod_l B_{J'_l,\,\Jhat'_l}
\left [{\varpi_0\over \Gamma\bigl(\varpihat_3 \bigr)
\Gamma\Bigl(1+\muhat_2+\nuhat_2+(1+\mu_2+\nu_2)h/\pi\Bigr)
\Gamma\bigl(\varpihat_1 \bigr)}\right ]^2.
\label{5.1}
\eeq
It immediately follows from Eq.\ref{4.21} that
\beq
e^{i\varphi(-\varpi_0,-\varpi_3)}
e^{i\varphi(-\varpi_3,\varpi_1)}
e^{i\varphi(\varpi_1,\varpi_0)}=1.
\label{5.2}
\eeq
One arrives at the completely symmetric
expression
\beq
{\cal C}_{1,2,3}=-\left ({\pi \over h} \Gamma\bigl(2+h/\pi
\bigr)
\Gamma\bigl (2-h/\pi \bigr ) \right )^2
\prod_l {B_{J'_l,\,\Jhat'_l}\over
\left [
\Gamma\Bigl(1+2\Jhat_l+(1+2J_l)h/\pi\Bigr)
\right  ]^2}.
\label{5.3}
\eeq

Next, we make contact with other approaches to the same problem.
For this we first explictly connect the present group-theoretic
notations with more standard conventions. Eqs.\ref{2.1}, and
\ref{3.2}, correspond to the usual notations $C=1+3Q^2$, and
$D=1-12\alpha_0^2$,  for the gravity and matter central charges.
The screening charges are given by (we choose $\alpha_0>0$)
\beq
\alpha_\pm ={Q\over 2}\pm \alpha_0,\quad
\alpha'_\pm=i(\alpha_0\mp{Q\over 2}).
\label{5.6}
\eeq
The dressed vertex operator Eq.\ref{3.10} may be rewritten as
\beq
{\cal V}_{J',\,\Jhat'}\equiv
 e^{\textstyle ((\Jhat'+1)\alpha_--J' \alpha_+)\Phi}
\>
e^{\textstyle -(J'\alpha'_-+\Jhat' \alpha'_+)
X} =e^{\textstyle \beta(k) \Phi-i
k X/\alpha_-},
\label{5.7}
\eeq
where
\beq
\beta=(\Jhat'+1)\alpha_--J' \alpha_+,\quad
ik=  \alpha_-(J'\alpha'_-+\Jhat'\alpha'_+).
\label{5.8}
\eeq
A simple calculation, using the formulae just given leads to
\beq
k=2J'-2\Jhat'h/\pi= 2\Jhat+2(J+1)h/\pi,
\label{5.9}
\eeq
\beq
\beta(k)+Q/ 2= (k+k_0)/\alpha_-,
\quad  k_0=\alpha_0 \alpha_-=1-h/\pi
\label{5.10}
\eeq
According to Eq.\ref{3.7}, the weights are given by
\beq
\Delta_{\cal G}=-{1\over 2} \beta (\beta+Q),\quad
\Delta_{\cal M}={\pi \over 4h } k(k+2k_0).
\label{5.11}
\eeq
For rational theories ($D=1-6(p-p')^2/pp'$, $p>p'>0$),
\beq
h=-\hhat',=\pi p'/ p,\quad
\hhat=-h',=\pi p/ p', \quad k_0=(p-p')/p,
\label{5.12}
\eeq
$\Delta_{\cal M}$ reduces to Kac's table
\beq
\Delta_{\cal M}={1\over 4pp'}
\left [ (rp-sp')^2-(p-p')^2 \right ],
\quad r=2J'+1,\> s=2\Jhat'+1,
\quad k+k_0=r-sh/\pi.
\label{5.13}
\eeq
The momentum $k$ is defined so that  it takes rational values
for rational theories. For critical bosonic string with
Regge slope $\alpha'$, the tachyon vertex is
$\exp (ikX\sqrt {\alpha'})$.  Eq.\ref{5.7} corresponds to
$\alpha'=1/(\alpha_-)^2$.
With the conventions just introduced,
the result Eq.\ref{5.3} takes the form
\beq
{{\cal C}_{1,2,3}\over -\left (\pi
\Gamma\bigl(2+h/\pi
\bigr)
\Gamma\bigl (2-h/\pi \bigr )/h  \right )^2} \equiv
{\cal A}(k_1,k_2,k_3)
=\prod_l {B(k_l)\over
\left [
\Gamma(k_l+k_0)
\right  ]^2},
\label{5.14}
\eeq
where we left out an  overall factor
 which does not depend
upon the momenta.

Our next task is to re-establish the cosmological constant. In the
present approach, it comes out as follows. Our  basic quideline
was to write down the most general local operators, as was
discussed in section 2.   We have not yet really done so, since we
may multiply the right-hand sides of Eq.\ref{2.19} (or more generally
of  Eq.\ref{2.37}) by
$\mu_c^{-\varpi/2}$ on the left, and by
$\mu_c^{\varpi/2}$ on the right,  without breaking
locality. This constant $\mu_c$ is arbitrary, and will play the
role of the cosmological constant. According to Eq.\ref{2.41} one has
\beq
 \mu_c^{-\varpi/2} \psi_{m\, \mhat}^{(J\, \Jhat )}
\>\mu_c^{\varpi/2}
= \mu_c^{m+\mhat \pi/h}
\psi_{m\, \mhat}^{(J\, \Jhat )}.
\label{5.15}
\eeq
In the Coulomb-gas picture, the power in $\mu_c$ is equal to the
total number of screening charges. We shall agree with this
definition if we let
\beq
e_{(\mu_c)}^{\textstyle -(J\alpha_-+\Jhat \alpha_+)
\Phi(\sigma, \tau )}=
 \mu_c^{J+\Jhat \pi /h}\> \mu_c^{-\varpi/2}
e^{\textstyle -(J\alpha_-+\Jhat \alpha_+)\Phi(\sigma, \tau )}
\>\mu_c^{\varpi/2}.
\label{5.16}
\eeq
Thus we have, according to Eq.\ref{3.10},
\beq
{\cal V}^{(\mu_c)}_{J',\,\Jhat'}(\sigma,\tau)=
\mu_c^{-(\Jhat'+1)+J' \pi /h} \mu_c^{-\varpi/2}
{\cal V}_{J',\,\Jhat'}(\sigma,\tau)
\>\mu_c^{\varpi/2}.
\label{5.17}
\eeq
The first factor coincides with the KPZ-DDK  scaling
factor\cite{KPZ,Da,DK}. In
particular, one has
\beq
{\cal V}^{(\mu_c)}_{0,\,0}(\sigma,\tau)=
\mu_c^{-1} \mu_c^{-\varpi/2}
{\cal V}_{0,\,0}(\sigma,\tau)
\>\mu_c^{\varpi/2},
\label{5.18}
\eeq
which is the expected scaling behaviour of the cosmological constant.
The operators   $\mu_c^{\pm\varpi/2}$ induce a translation on the
variable $x$ conjugate to the Liouville momentum. This  degree of
freedom already  appears in the double free field representation
recalled in appendix A. With the rescaled momentum $\varpi$, the
natural definition of x is such that $[x,\varpi]=i$, and
$\psi_{m\, \mhat}^{(J\, \Jhat )}$
is proportional to $\exp [-2ix(m+\mhat\pi/h)]$,
in agreement with Eq.\ref{5.15}. This Liouville position-operator is
equal to $-i\sqrt{h/2\pi} q_0^{(1)}$, where $q_0^{(1)}$ is the zero
mode of the free field $\phi_1$ whose properties are summarized
in the appendix.
 This translation,   which is actually a global Weyl
transformation,  is analogous  to
the translation of the Liouville field in the work of  DDK.
This is seen by computing next the $\mu_c$-dependence of the
three-point function. One gets immediately
\beq
{\cal A}_{(\mu_c)}(k_1,k_2,k_3)=
{\cal A}(k_1,k_2,k_3)
\> \mu_c^{\varpi_0+\sum_l(J_l+\Jhat_l\pi/h)}.
\label{5.19}
\eeq
The term $\mu_c^{\varpi_0}$ arises when
$\mu_c^{-\varpi/2}$, and $\mu_c^{\varpi/2}$ hit the
left and right vacua respectively. In the DDK
discussion\cite{Da,DK}, it  comes from the term of the
effective action which is  linear
 in the field\footnote{ It would  not be there on
the torus, since
one would take a trace. This agrees with the DDK result where its
contribution is shown to be   proportional to the Euler characteristic.}.
According to
Eq.\ref{3.47}, the power of $\mu_c$ is equal to   $\nu_2+\nuhat_2\pi/h$
where $\nu_2$ and $\nuhat_2$ are the gravity screening-numbers.
This agrees with the usual definition (see, e.g. ref.\cite {D}).

Finally we compare Eq.\ref{5.14} with the result of the matrix model. This
part follows ref.\cite{D} closely.
 The
two-point function ${\cal A}_{(\mu_c)}(k,k)$ is determined
by  starting from the
three-point function with one cosmological term
${\cal A}_{(\mu_c)}(0,k,k)$, and writing
\beq
{d\over d\mu_c} {\cal A}_{(\mu_c)}(k,k)
={\cal A}_{(\mu_c)}(0,k,k).
\label{5.20}
\eeq
According to Eq.\ref{5.18}, this gives
\beq
 {\cal A}_{(\mu_c)}(k,k)
={\mu_c \over k+k_0} {\cal A}_{(\mu_c)}(0,k,k).
\label{5.21}
\eeq
Similarly,  the partition function satisfies
\beq
{d^3\over d\mu_c^3} {\cal Z}_{(\mu_c)}
={\cal A}_{(\mu_c)}(0,0,0),
\label{5.22}
\eeq
\beq
 {\cal Z}_{(\mu_c)}={(\mu_c)^3\over (\varpi_0) (\varpi_0-1)
(\varpi_0-2)} {\cal A}_{(\mu_c)}(0,0,0).
\label{5.23}
\eeq
We finally obtain the rescaled three point function
\beq
{\cal D}(k_1,k_2,k_3) \equiv
{\left ({\cal A}_{(\mu_c)}(k_1,k_2,k_3)\right )^2 {\cal
Z}_{(\mu_c)}\over \prod_l {\cal A}_{(\mu_c)}(k_l,k_l)}
={\prod_l (k_l+k_0)\over (\pi/h+1)(\pi/h)(\pi/h-1)},
\label{5.24}
\eeq
which agrees with the results of the matrix models.
Clearly, the key point in this final verification is that the
final expression Eq.\ref{5.14} for the three-point function factorises.
On the other hand, Eq.\ref{5.14}  vanishes whenever
$k_l+k_0=0$ for any of the
three legs. According to Eq.\ref{5.13}, this happens
for  rational theories,
at the border of Kac's table, where $r=p'$, and $s=p$. Thus formula
Eq.\ref{5.24} holds only when the branching rules are satisfied.

\section{ CONCLUSION}
\markboth{6.  Conclusion} {6.  Conclusion}
\label{6}

Our starting point was    the operator expressions of the
Liouville and matter fields which are
derived by imposing  locality and consistency of the  restriction to the
Hilbert space of states with
equal left and right momenta. These requirements uniquely determine
these fields, the only arbitrariness being the
choice of   cosmological constant.
We have seen how the three-point functions
  tremendously simplify when gravity and
matter are coupled, so that the matrix-model results on the sphere
come  out.

The present  operator viewpoint has given us an insight into the treatment
of negative spins (positive powers of the metric)
 necessary for the proper dressing by gravity. Two kinds of operators
of this type were found to appear. On the one hand,
the symmetry between spin $J$ and spin $-J-1$ showed the existence of
the local
 field  $\exp \left ((J+1)\alpha_- \widetilde \Phi\right )$. It is
similar to
$\exp \left (-J \alpha_-   \Phi\right )$, since its expansion
has an index running from $-J$ to $J$. It is appropriate for
matrix elements between on-shell physical states. On the other hand,
the treatment  of the end points, requires the use of a field which is
really the quantum analogue of the classical expression
 Eq.\ref{1.14}.
 In spite of the  lack of explicit symmetry in our
definition of the three-point functions, the three legs play the
same role in our final formulae. This situation is typical of a
picture changing mechanism as the one of critical
superstrings\cite{GSW}.  It might be possible to redefine the
Fock spaces so that the complete symmetry becomes manifest.
The situation is similar to the Neveu-Schwarz model in the old
formulation, with the wrong vacuum. The possible change of picture is
left
for further studies.

In any
case,
since ${\cal C}_{1,2,3}=
{\cal C}_{2,1,3}$, and ${\cal C}_{1,2,3}=
{\cal C}_{1,3,2}$, it follows that
\beqa
 <-\varpi_3, -\varpi'_3\vert  {\cal V}_
{J'_2,\,\Jhat'_2}(\sigma,\,\tau)
 {\tilde {\cal V}}_
{J'_1,\,\Jhat'_1} (\sigma',\,\tau) \vert  \varpi_0, \varpi'_0 >=
\nnn
<-\varpi_3, -\varpi'_3\vert  {\cal V}_
{J'_1,\,\Jhat'_1}(\sigma',\,\tau)
 {\tilde {\cal V}}_
{J'_2,\,\Jhat'_2} (\sigma,\,\tau) \vert  \varpi_0, \varpi'_0 >
\label{6.1}
\eeqa
\beqa
<-\varpi_0, -\varpi'_0\vert  {\tilde {\cal V}}_
{J'_3,\,\Jhat'_3}(\sigma,\,\tau)
 {\cal V}_
{J'_2,\,\Jhat'_2} (\sigma',\,\tau)  \vert  \varpi_1, \varpi'_1 >
\nnn
=<-\varpi_0, -\varpi'_0\vert  {\tilde {\cal V}}_
{J'_2,\,\Jhat'_2}(\sigma',\,\tau)
 {\cal V}_
{J'_3,\,\Jhat'_3}(\sigma,\,\tau)  \vert  \varpi_1, \varpi'_1 >.
\label{6.2}
\eeqa
Thus the fields ${\cal V}$ and ${\tilde {\cal V}}$ are
mutually local in a generalised sens.

It is rather remarkable that the symmetry between spins  $J$ and $-J-1$
precisely gave us the operator $\widetilde \Phi$ which
 is needed to treat the middle point, that is, to define the emission
operators between physical states. For its exponentials,
Eq.\ref{3.17}
 only involve a
finite number of terms, contrary to  what one could expect in view
of the classical structure. In particular,  the cosmological term
$\exp \left (\alpha_- \widetilde \Phi\right )$
corresponds to  $J=0$, so that
it only involves   a single  term on the right-hand side! This is in sharp
contrast with the direct quantization of the classical cosmological
term which has an expansion over an infinite number of conformal
blocks with
different screening charges. This tremendous simplification may be the
deep reason why matrix models give results to  all orders in
perturbation.

Since the present operator formalism is not restricted to the
sphere, higher genus may certainly be considered. The case of the
torus,  should be straighforward.

Many future developments of the present discussion may be forseen.
In particular, the present quantum-group
approach  is, so far,  the only one that has been able to
overcome the $D=1$ barrier\cite{G2,G3}. I hope to return to this
in latter publications.
\section { Acknowledgements}
This work was started when I was visiting the {\sl Mathematical
Sciences
Research Institute}, in Berkeley, California, during March 1991.
It is a pleasure to acknowledge the  generous financial support, the
very stimulating surrounding and the warm hospitality from which
I benefitted very much  during my participation in the program
``Strings in Mathematical Physics''.

\appendix
\section{Appendix}
\markboth{Appendix}{Appendix}

   In this appendix we recall the basic properties of the
$x_+$
components which are
holomorphic functions of $z=\tau+i\sigma$.
 Since one deals with functions
of a single  variable, one may work at $\tau=0$
without loss of generality
that is on the unit circle $u=e^{i\sigma}$.
The starting point is that,  for generic $\gamma$,
there exist two equivalent free fields:
\beq
\phi_j(\sigma)=q^{(j)}_0+ p^{(j)}_0\sigma+i
\sum_{n\not= 0}e^{-in\sigma}\, p_n^{(j)}\bigl / n,
\quad j=1,\> 2,
\label{A.1}
\eeq
such that
\beq
\Bigl [\phi'_1(\sigma_1),\phi'_1(\sigma_2) \Bigr ]=
\Bigl[\phi'_2(\sigma_1),\phi'_2(\sigma_2) \Bigr ]
=2\pi i\,  \delta'(\sigma_1-\sigma_2),
\qquad p_0^{(1)}=-p_0^{(2)},
\label{A.2}
\eeq
\beq
N^{(1)}\bigl (\phi'_1\bigr )^2+ \phi''_1/\sqrt \gamma=
N^{(2)}\bigl ( \phi'_2\bigr)^2+ \phi''_2/\sqrt \gamma.
\label{A.3}
\eeq
$N^{(1)}$ (resp. $N^{(2)}$)  denote  normal orderings
 are with respect to the modes of $\phi_1$  (resp. of
$\phi_2$).
 Eq.\ref{A.3} defines the
stress-energy tensor  and the  coupling constant
$\gamma$ of  the quantum theory. The former  generates a
representation of the Virasoro algebra
with central charge $C=3+1/\gamma$. At an intuitive
level, the
correspondence between $\phi_1$ and $\phi_2$ may be
understood
from the fact that the Verma modules, which they  generate,  coincide
since   the highest weights only depend upon
$(p_0^{(1)})^2=(p_0^{(2)})^2$.
The chiral family is built up\cite{GN4,GN5,GN7,G1,Gr} from
the
 following operators
\beq
\psi_j =d_j\, N^{(j)}\bigl (e^{\sqrt{h /2\pi}\>
\phi_j}\bigr
), \quad
\psihat_j ={\widehat d}_j\,
N^{(j)}\bigl (e^{\sqrt{\hhat /2\pi}\> \phi_j}\bigr ),
\quad
j=1,\>2,
\label{A.4}
\eeq
\beq
h={\pi \over 12}\Bigl(C-13 -
\sqrt {(C-25)(C-1)}\Bigr),\quad
\hhat={\pi \over 12}\Bigl(C-13
+\sqrt {(C-25)(C-1)}\Bigr),
\label{A.5}
\eeq
where $d_j$ and $\dhat_j$ are normalization constants.
In Eq.\ref{A.4}, the zero modes are ordered, as is
standard in string vertices, so that the operators are
primary.
Explicitly one has
$$
  N^{(j)}\bigl( e^{\sqrt{h/2\pi}\, \phi_j} \bigr)
\equiv
e^{\sqrt{h/2\pi}\, q_0^{(j)}}\,e^{ \sqrt{h/2\pi}
p_0^{(j)}}
e^{-i\sigma h/4\pi}\times
$$
\beq
 \exp\left (\sqrt{h/2\pi}i\sum_{n<0}e^{-in\sigma}p_n^{(j)}/ n\right )
\exp\left (\sqrt{h/2\pi}i\sum_{n>0}e^{-in\sigma}p_n^{(j)}/ n\right )
\label{A.6}
\eeq
with similar definitions for the hatted fields.
The relation between  $h$ or $\hhat$ and $C$ which is
equivalent
to
\beq
C=1+6({h\over\pi}+{\pi\over h}+2)=
1+6({\hhat\over\pi}+{\pi\over\hhat}+2),
\quad \hbox{with} \quad h\hhat=\pi^2,
\label{A.7}
\eeq
is such that $\psi_j$ and $\psihat_j$ are  solutions of
the
 equations\cite{GN3,GN4}
$$
 -{d^2\psi_j(\sigma)\over d\sigma^2}+({h\over
\pi})\Bigl(\sum_{n<0}\,L_n\,e^{-in\sigma}
+{L_0\over 2}+({h\over 16\pi}-{C-1\over
24})\Bigr)\psi_j(\sigma)
$$
\beq
+({h\over
\pi})\psi_j(\sigma)\Bigl(\sum_{n>0}\,L_n\,e^{-in\sigma}
+{L_0\over 2}\Bigr)=0
\label{A.8}
\eeq
$$ -{d^2\psihat_j(\sigma)\over d\sigma^2}+({\hhat\over
\pi})
\Bigl(\sum_{n<0}\,L_n\,e^{-in\sigma}
+{L_0\over 2}+({h\over 16\pi}-{C-1\over
24})\Bigr)\psihat_j(\sigma)
$$
\beq
+({\hhat\over
\pi})\psihat_j(\sigma)\Bigl(\sum_{n>0}\,L_n\,e^{-in\sigma}
+{L_0\over 2}\Bigr)=0
\label{A.9}
\eeq
These are  operator Schr\"odinger equations equivalent
to the decoupling of Virasoro
null-vectors\cite{GN4,GN5,GN7}.
Since there are two possible quantum modifications $h$
and
$\hhat$,
there are four solutions.
By operator product $\psi_j$, $j=1$, $2$, and $\psihat
_j$,
$j=1$, $2$,
generate two infinite families of chiral
fields  $\psi_m^{(J)}$, $-J\leq m \leq J$,
and
$\psihat_\mhat^{(\Jhat)}$, $-\Jhat\leq \mhat \leq \Jhat$;
  with $\psi_{-1/2}^{(1/2)}= \psi_1$,
$\psi_{1/2}^{(1/2)}= \psi_2$, and
$\psihat_{-1/2}^{(1/2)}=
\psihat_1$ ,
$\psihat_{1/2}^{(1/2)}= \psihat_2$.
The fields
$\psi_m^{(J)}$, $\psihat_\mhat^{(\Jhat)}$,  are
 of the type ($1$, $2J+1$) and ( $2\Jhat+1$,$1$),
respectively,
in the BPZ
classification. For the zero-modes, it is
simpler\cite{G1}
to  define  the rescaled variables
\beq
\varpi=i p_0^{(1)} \sqrt{{2\pi\over h }}; \quad
\varpihat=i p_0^{(1)} \sqrt{{2\pi \over \hhat }};
\qquad \varpihat=\varpi\>{h\over \pi};
\quad \varpi=\varpihat \>{\hhat\over \pi}.
\label{A.10}
\eeq
The Hilbert space in which the operators $\psi$ and
$\psihat$
live,
 is a direct sum\cite{G1,GR,G2,G3,Gr} of Fock spaces
${\cal F}(\varpi)$ spanned by the harmonic excitations of
 highest-weight Virasoro states noted $\vert \varpi>$.
They  are eigenstates of the quasi momentum $\varpi$, and
satisfy $L_n\vert \varpi>= 0$, $n>0$;
$(L_0~-~\Delta(\varpi))\vert \varpi>~=~0$.
The corresponding highest weights  $\Delta (\varpi)$
 may be rewritten as
\beq
\Delta(\varpi)\equiv {1\over
8\gamma}+{(p_0^{(1)})^2\over 2}
={h\over 4\pi}(1+{\pi\over h})^2-{h\over 4\pi}\varpi^2.
\label{A.11}
\eeq
The commutation relations Eq.\ref{A.2} are to be supplemented
by the
zero-mode
ones:
$$
 \bigl[ q^{(1)}_0,\, p^{(1)}_0\bigr]=
\bigl[ q^{(2)}_0,\, p^{(2)}_0\bigr]=i.
$$
It thus follows (see in particular Eq.\ref{A.6}), that the
fields $\psi$ and $\psihat$  shift the quasi momentum
$p^{(1)}_0=-p^{(2)}_0$ by a fixed amount.
For an arbitrary c-number  function  $f$ one has
\beq
\psi_m^{(J)}\>f(\varpi)=f(\varpi+2m)\>\psi_m^{(J)},\quad
 \psihat_\mhat^{(\Jhat)}\>f(\varpi)=
f(\varpi+2\mhat\, \pi/h)\>\psihat_\mhat^{(\Jhat)}.
\label{A.12}
\eeq
The fields $\psi$ and $\psihat$ together with
their products may be naturally restricted to discrete
values
of $\varpi$. They thus live    in Hilbert
spaces\footnote{
 Mathematically they
are not really Hilbert spaces since
their metrics are not  positive definite.}  of the form
\beq
{\cal H}(\varpi_0)\equiv
\bigoplus_{n,\nhat=-\infty}^{+\infty}
{\cal F}(\varpi_0+n+\nhat\,\pi/h).
\label{A.13}
\eeq
 $\varpi^0$ is a constant which is arbitrary so far. The
$sl(2,C)$--invariant
vacuum corresponds to $\varpi_0=1+\pi/h$,\cite{G1}. For
$h$
real, it is thus natural that the eigenvalues of $\varpi$
be
real. Thus $(p^{(1)}_0)^\dagger =p^{(2)}_0$. As discussed
several times\cite{G1}, the natural hermiticity relation
is
$\bigl(\phi^{(1)}(\sigma)\bigr)^\dagger
=\phi^{(2)}(\sigma)$.
Eq.\ref{A.3} shows that this is consistent with the usual
hermiticity relation $L_n^\dagger=L_{-n}$. As shown in
\cite{G1}, one has\footnote{ There are subtleties
which we are discussed in section 3. }
\beq
(\psi_m^{(J)})^\dagger=\psi_{-m}^{(J)}
\label{A.14}
\eeq
 For latter reference, we
note that,
besides their simple branching rules between Fock spaces,
  the $\psi$ fields enjoy
another important property: they are periodic up
to a constant. This may be seen as follows: First it is
obvious
from the definitions  Eqs.\ref{A.4}, \ref{A.6}, \ref{A.10}   that
\beq
\psi_{\pm 1/2}^{(1/2)}(\sigma+2\pi)=e^{\pm ih\varpi}
e^{ih/2}
\, \psi_{\pm 1/2}^{(1/2)}(\sigma)
\label{A.15}
\eeq
Next this is extended to the other $\psi$ fields by
fusion.
To leading order at $\sigma_1\to \sigma_2$ one
has\cite{G1}
\beq
 \psi_{\pm 1}^{(1/2)}(\sigma_1)
\psi_m^{(J)}(\sigma_2) \sim
(1-e^{-i(\sigma_1-\sigma_2)})^{-hJ
/ \pi}\> {\sin [h( \varpi \mp J +m)] \over \sin h}
 \psi_{m+\alpha}^{(J+1/2)}(\sigma_1),
\label{A.16}
\eeq
from which is follows that the fields $\psi_m^{(J)}$
satisfy
\beq
\psi_m^{(J)}(\sigma+2\pi ) =  e^{2ihm\varpi}
e^{2ihm^2}\,
\psi_m^{(J)}(\sigma)
\label{A.17}
\eeq
The fields $\psi_m^{(J)}$ are quantum Bloch waves. The
second
term in Eq.\ref{A.17} is a quantum modification.  Its role is
easily seen when one checks that Eq.\ref{A.17}  is compatible
with
the hermiticity condition Eq.\ref{A.14}, if one takes Eq.\ref{A.12}
into
account.  The properties
of the fields $\psihat$ are of course similar, and we
need not
elaborate upon them.

An important point is that the
$U_q(sl(2))$-quantum-group
structure is best seen  by changing basis
to new chiral operators  which obey a much simpler
   operator algebra. Following my recent work,\cite{G1}
let us
introduce
\beq
\xi_M^{(J)}(\sigma) := \sum_{-J\leq m \leq J}\vert
J,\varpi)_M^m \>
\psi_m^{(J)}(\sigma),  \quad -J\leq M\leq J;
\label{A.18}
\eeq
\beqa
&\vert J,\varpi)_M^m\,=
\sqrt{\hbox{$ {2J \choose J+M} $ }} \>e^{ihm/2}\times \nnn
&\sum_{(J-M+m-t)/2 \>\>\hbox{integer}}\, e^{iht(\varpi
+m)}
 {J-M \choose (J-M+m-t)/2}\, {J+M \choose (J+M+m+t)/2};
\label{A.19}
\eeqa
\beq
 {P \choose Q} \equiv {\lfloor P \rfloor \! !
 \over \lfloor Q \rfloor \! !\lfloor P-Q \rfloor \! !}
\quad \lfloor n \rfloor \! ! \equiv
\prod_{r=1}^n \lfloor r \rfloor
\quad \lfloor r \rfloor \equiv {\sin (hr)\over \sin h}.
\label{A.20}
\eeq
The last equation introduces q--deformed factorials and
binomial coefficients. The other fields
$\xihat_\Mhat^{(\Jhat)}$
are defined in exactly the same way replacing $h$ by
$\hhat$
everywhere.
The symbols are the same with hats, e.g.
\beq
\lfloorhat n \rfloorhat \! ! \equiv
\prod_{r=1}^n \lfloorhat r \rfloorhat,
\qquad \lfloorhat r \rfloorhat \equiv {\sin (\hhat
r)\over \sin
\hhat},
\qquad \hbox{ and so on.}
\label{A.21}
\eeq
The above transformation may be explicitly inverted
\cite{G3}.
 One
has
\beq
\psi_m^{(J)}(\sigma) =\sum_{M=-J}^J
\xi_M^{(J)}(\sigma)\> (J, \varpi|_m^M
\label{A.22a}
\eeq
\beq
(J, \varpi|_m^M
=  \left (2i \sin h e^{ih/2}\right )^{2J}
(-1)^{J+M}\,e^{ih(J+M)}\,
\vert J,\,\varpi )_{-M}^{-m} {\lfloor \varpi-2m \rfloor
(-1)^{J+m}
\over \lambda_{-m}^{(J)}(\varpi)}
\label{A.22}
\eeq
\beq
\lambda_m^{(J)}(\varpi):=
{2J \choose J-m}
\lfloor \varpi-J+m \rfloor_{2J+1}.
\label{A.23}
\eeq
In general, we define
\beq
\lfloor x \rfloor_{N} =
\lfloor x \rfloor \lfloor x+1 \rfloor \cdots
\lfloor x+N-1 \rfloor.
\label{A.24}
\eeq
In \cite{G1,G3} the operator-algebra of the  $\xi$
fields
was completely determined\footnote{The
fusion  propertes were  not really proven  so far, but rather made
very plausible from quantum group invariance\cite{G3}. Its  complete
derivation will be given elsewhere\cite{CGR}.}. In particular, it was shown
that
for $\pi>\sigma_1>\sigma_2>0$,
these operators obey the exchange algebra
\beq
\xi_{M_1}^{(J_1)}(\sigma_1)\,\xi_{M_2}^{(J_2)}(\sigma_2)=
\sum_{-J_1\leq N_1\leq J_1;\> -J_2\leq N_2\leq J_2}\>
(J_1,J_2)_{M_1\, M_2}^{N_2\, N_1}\, \xi_{N_2}^{(J_2)}
(\sigma_2) \,\xi_{N_1}^{(J_1)}(\sigma_1),
\label{A.25}
\eeq
$$( J_1,J_2)_{M_1\,M_2}^{N_2\,N_1}=
\delta(M_1+M_2-N_1-N_2)
\,e^{-2ih M_1M_2}\>(1-e^{2ih })^n\times
$$
\beq
 {e^{ih n(n-1)/2}\over \lfloor n \rfloor \! !} \,
e^{-ih n(M_1-M_2)}\sqrt{ {
\lfloor J_1+M_1 \rfloor \! ! \,
\lfloor J_1-N_1 \rfloor \! ! \,
\lfloor J_2-M_2 \rfloor \! ! \,
\lfloor J_2+N_2 \rfloor \! ! \over
\lfloor J_1-M_1 \rfloor \! ! \,
\lfloor J_1+N_1 \rfloor \! ! \,
\lfloor J_2+M_2 \rfloor \! ! \,
\lfloor J_2-N_2 \rfloor \! !}},
\label{A.26}
\eeq
where $n=M_1-N_1=N_2-M_2$. For $\pi>\sigma_2>\sigma_1>0$,
on the other hand, one has
\beq
\xi_{M_1}^{(J_1)}(\sigma_1)\,\xi_{M_2}^{(J_2)}(\sigma_2)=
\sum_{-J_1\leq N_1\leq J_1;\> -J_2\leq N_2\leq J_2}\>
{\overline {(J_1,J_2)}_{M_1\, M_2}^{N_2\, N_1}}\,
\xi_{N_2}^{(J_2)}
(\sigma_2) \,\xi_{N_1}^{(J_1)}(\sigma_1),
\label{A.27}
\eeq
$${\overline {(J_1,J_2)}}_{M_1\,M_2}^{N_2\,N_1}=
\delta(M_1+M_2-N_1-N_2)
\,e^{2ihM_1M_2} (1-e^{-2ih})^m \times$$
\beq
{e^{-ihm(m-1)/2}\over \lfloor m \rfloor \! !}
\,
e^{ihm(M_2-M_1)}\sqrt{ {
\lfloor J_1-M_1 \rfloor \! ! \,
\lfloor J_1+N_1 \rfloor \! ! \,
\lfloor J_2+M_2 \rfloor \! ! \,
\lfloor J_2-N_2 \rfloor \! ! \over
\lfloor J_1+M_1 \rfloor \! ! \,
\lfloor J_1-N_1 \rfloor \! ! \,
\lfloor J_2-M_2 \rfloor \! ! \,
\lfloor J_2+N_2 \rfloor \! !}},
\label{A.28}
\eeq
where $m=M_2-N_2=N_1-M_1$.
The two braiding matrices are related by the inverse
relation
\beq
\sum_{-J_1\leq N_1 \leq J_1;\> -J_2\leq N_2 \leq J_2}
 (J_1,J_2)_{M_1\, M_2}^{N_2\, N_1}\>
{\overline {(J_2,J_1)}_{N_2\, N_1}^{P_1\,P_2}}
=\delta_{M_1,P_1}\,\delta_{M_2,P_2}.
\label{A.29}
\eeq
$ ( J_1,J_2)_{M_1\,M_2}^{N_2\,N_1}$ is given\cite{G1} by
the
corresponding matrix element of the  universal R matrix
of
$U_q(sl(2))$. Concerning  the hermiticity of the $\xi$
fields,
it was shown in \cite{G1} that, for $\varpi$ real,
\beq
 \bigl (\vert J,\omega+2m)_M^{-m}\bigr)^\dagger=\vert
J,\omega)_M^{m}
\label{A.30}
\eeq
so that
\beq
\xi_M^{(J)}(\sigma)^*=\xi_M^{(J)}(\sigma)
\label{A.31}
\eeq
is a  hermitian field.
 Obviously the same structure holds for the hatted
fields. One
replaces
$h$ by $\hhat$ everywhere. Moreover the hatted and
unhatted
fields have simple braiding and fusions\cite{G1}.  The
most
general
$(2\Jhat+1, 2J+1)$
field $\xi_{M\,\Mhat}^{(J\,\Jhat)}\sim
\xi_{M}^{(J)}\>\xihat_{\Mhat}^{(\Jhat)}$ has weight
\beq
 \Delta_{Kac} (J,\Jhat;C)={C-1\over 24}-
{ 1 \over 24} \left((J+\Jhat+1) \sqrt{C-1}
-(J-\Jhat) \sqrt{C-25} \right)^2,
\label{A.32}
\eeq
in agreement with Kac's formula.

\typeout{========================================================}
\typeout{ This was the Latex file of the article:}
\typeout{      ``GRAVITY-MATTER COUPLINGS FROM LIOUVILLE THEORY''}
\typeout{                   by Jean-Loup GERVAIS}
\typeout{(Please run twice to get the references right)}
\typeout{========================================================}


\markboth{References}{References}
\begin{thebibliography}{99}

\bibitem{GL} M. Goulian, M. Li, \prl 66, 2051, 1991.

\bibitem{D} Vl.S. Dotsenko, \mpl A6,  3601, 1991.

\bibitem{Coulomb} P. di Francesco, D. Kutasov,
\pl B261, 385, 1991;  Y. Kitazawa, \pl B265, 262, 1991;
N. Sakai, Y. Tanii, \ptp 86, 547, 1991;
L. Alvarez-Gaum\'e, J.L. Barb\'on, C. G\'omez,
preprint SLAC-PUB-CERN-TH.6142/91, June 1991;
K. Aoki, E. D'Hoker, preprint UCLA/91/TEP/32,
August 1991.

\bibitem{B} O. Babelon,
\pl   B215,  523, 1988.

\bibitem{G1} J.-L. Gervais,  \cmp 130, 257, 1990.

\bibitem{G2} J.-L. Gervais, \pl B243, 85, 1990.

\bibitem{G3} J.-L. Gervais, \cmp 138, 301, 1991.

\bibitem{G4} J.-L. Gervais, ``On the algebraic structure
of
Quantum gravity in two dimensions'',
  Proceedings of the
Trieste Conference on  {\sl Topological Methods in Quantum Field
Theories}, June 1990, W. Nahm, S. Randjbar-Daemi, E.
Sezgin, E. Witten editors, World Scientific.

\bibitem{GR} J.-L. Ger\-vais, B. Ros\-tand,
 \cmp 143, 175, 1991.

\bibitem{CG} E. Cremmer, J.-L. Gervais,
 \cmp 144, 279, 1992.

\bibitem{G5} J.-L. Gervais, ``Quantum group derivation
of 2D gravity-matter coupling'' Invited talk at
the Stony Brook meeting {\sl String and Symmetry 1991}.

\bibitem{Da} F. David, {\sl C. R. Acad. Sci. Paris} {\bf
307},
1051,1988; {\sl Mod. Phys. Lett. A} {\bf 17}, 1651, 1988.

\bibitem{DK} J. Distler, H. Kawai, \np B321, 509, 1651.

\bibitem{B2} O. Babelon, ``Universal exchange algebra
for Bloch waves and Liouville theory'' preprint
PAR LPTHE 91/11.

\bibitem{GN1} J.-L. Gervais, A. Neveu,  \np B199, 59,
1982.

\bibitem{GN2} J.-L. Gervais, A. Neveu,  \np B202, 125,
1982.

\bibitem{GN4} J.-L. Gervais, A. Neveu \np B238, 125,
1984;
\np B238, 396, 1984.


\bibitem{GN6} J.-L. Gervais, A. Neveu, \pl 151B, 271,
1985.

\bibitem{CGR} E. Cremmer, J.-L. Gervais, J.-F. Roussel,
to be published.

\bibitem{GS} J.-L. Ger\-vais, B. Sakita, \pr  D4, 2291,
1971.

\bibitem{BP} J. Balog, L. Palla, \pl B274, 232, 1992.

\bibitem{KPZ} V. Knizhnik, A. Polyakov, A.A.
Zamolodchikov,
{\sl Mod. Phys. Lett.} {\bf A3}, 819, 1988.


\bibitem{GSW} M. Green. J. Schwarz, E. Witten
``Superstring theory'' {\sl Cambridge Monographs on
Mathematical Physics}  1987.

\bibitem{GN3} J.-L. Gervais, A. Neveu,  \np B224, 329,
1983.


\bibitem{GN5} J.-L. Gervais, A. Neveu,  \np B257[FS14], 59,
1985.


\bibitem{GN7} J.-L. Gervais, A. Neveu, \np B264, 557, 1986.

\bibitem{Gr} For reviews see,
J.-L. Gervais, ``Liouville Superstrings'' in
``Perspectives
in string the
proceedings of the Niels Bohr/Nordita Meeting (1987)
World
Scientific; DST
workshop on particle physics-Superstring theory,
proceedings
of the
 I.I.T. Kanpur meeting (1987) World Scientific;
J.-L. Gervais, `` Systematic approach
 to conformal theories'', {\sl Nucl. Phys. B
(Proc. Supp.)} {\bf 5B}, 119-136 (1988) 119;
A. Bilal, J.-L. Gervais, `` Conformal theories with
non-linearly-extended
Virasoro symmetries and Lie-algebra classification'',
Conference
 Proceedings: "Infinite dimensional Lie algebras and
Lie groups", edited by V. Kac, Marseille 1988,
World-Scientific.



\end{thebibliography}
\end{document}